\documentclass{aa} 
\usepackage{txfonts} 
\usepackage{graphicx}
\usepackage[breaklinks]{hyperref}
\begin{document} 
\title {The secrets of T Pyx II.\\ A recurrent nova that will not
become a
SN
Ia} 
\author{P. Selvelli \inst{1}\and A.
Cassatella
\inst{2}  R. Gilmozzi \inst{3} \and R. Gonz\'alez-Riestra \inst{4}}
\offprints{P. Selvelli}
\institute {INAF-Osservatorio
Astronomico di Trieste, Via Tiepolo 11 - Trieste, I-34143 Trieste,
Italy
 \and INAF-IFSI, Via del
Fosso
del Cavaliere 100, 00133 Roma, Italy,  and Dipartimento di Fisica,
Universita' Roma Tre, 00146 Roma, Italy
\and European Southern
Observatory,
 Karl-Schwarzschild-Str 2, D-85748 Garching bei M\"unchen,
Germany 
\and  XMM-Newton Science Operations Centre, ESAC, P.O. Box 78,
28691
Villanueva de la Ca\~nada, Madrid (Spain) } 
\date{Received.....; accepted }
 \abstract {} {
We compare  the observed and  theoretical
parameters for the quiescent and outburst phases of the recurring
nova T
Pyx.
} {
IUE data were used to derive the disk luminosity and the mass
accretion rate, and to exclude the presence of quasi-steady burning
at the
WD surface. XMM-NEWTON data were used to verify this 
conclusion.
} {
By various methods, we  obtained   L$_{disk}$
$\sim$ 70 L$_{\odot}$ and $\dot{M}$ $\sim$1.1 $\times$ 10$^{-8}$
M$_{\odot}$yr$^{-1}$. These values were about twice as high in the 
pre-1966-outburst epoch. This allowed the first direct estimate 
of the
total
mass accreted 
before outburst, M$_{accr}$=$\dot{M}_{pre-OB}$  $\cdot \Delta$t, 
and its
comparison  with
the critical ignition mass M$_{ign}$.  We found M$_{accr}$ and 
M$_{ign}$ to be 
in perfect  agreement (with a value close to 5 $\times$
10$^{-7}$M$_{\odot}$)
for M$_1$
$\sim$ 1.37 M$_{\odot}$,  which provides a confirmation of the
thermonuclear 
runaway
theory. 
 The comparison of the observed parameters of the eruption phase, with the
corresponding values in the grid of models by Yaron and collaborators,
provides satisfactory agreement for
values of M$_1$ close to 1.35 M$_{\odot}$ and log$\dot{M}$ between -8.0 and
-7.0,
but the observed  value of the decay time t$_3$ is higher than expected.
The long
duration of
the optically thick phase  during the recorded  outbursts of T Pyx, a
spectroscopic behavior typical of classical novae,  and the 
persistence of
P Cyg profiles,  constrains the   ejected mass M$_{ign}$ to within 
10$^{-5}$ -
10$^{-4}$ M$_{\odot}$. Therefore, T Pyx ejects far more material
than it
has  accreted, and the mass of the white dwarf will not increase to the
Chandrasekhar limit as generally believed  in recurrent novae.
A detailed study based on the UV data excludes the possibility that T Pyx
belongs
to the
class of the supersoft X-ray sources, as has been postulated.
XMM-NEWTON observations have revealed a weak, hard source and
confirmed
this interpretation. } {}  
\titlerunning{The secrets of T Pyx} \maketitle %

\section{Introduction} Classical novae (CNe) are semi-detached
binary
systems in which a white dwarf (WD) primary star accretes
hydrogen-rich
matter,
by means of an accretion disk, from a companion star which fills its  Roche
lobe.
The ``classical nova" outburst  is a thermonuclear runaway (TNR) process on
the
surface of a white dwarf that is produced when, due to the gradual
accumulation of hydrogen-rich material on its surface, the pressure
at the
bottom of the accreted layer, which is partially or fully
degenerate,
becomes sufficiently high (P $\ge$ 10$^{18}$-10$^{20}$ dyn
cm$^{-2}$ ) for
nuclear
ignition of hydrogen to begin. The critical mass for ignition
M$_{ign}$
depends
primarily on the mass of the white dwarf as M$_{ign}$ 
$\sim$ M$_1^{-7/3}$,
although more
detailed models show  that the ignition mass also depends 
on both the
mass accretion rate $\dot{M}$ and  the core WD temperature
(Prialnik and
Kovetz
1995; Townsley and Bildsten 2004;  Yaron et al. 2005; see also Sect.
8). 

Recurrent novae (RNe) are a subclass of classical novae 
characterized
by outbursts with
recurrence time of the order of decades. We refer to Webbink et al.
(hereinafter WLTO, 1987) as  the seminal paper on the nature
of
recurrent novae and
to Warner (1995), Hachisu and Kato (2001), and  Shore (2008) for 
further
considerations of this topic. Solid theoretical considerations 
indicate
that a model of  a recurrent outburst
requires a high accretion rate  $\dot{M}$ (10$^{-8}$-10$^{-7}$ M$_{\odot}$)
onto a
WD of
mass
close to the
Chandrasekhar mass limit (Starrfield 1985; WLTO 1987; Livio 1994).
However,
nova
models (Prialnik and Kovetz 1995; Yaron et al. 2005; see also Sect.
8) appear to evade
these
strict requirements and allow the occurrence  of recurrent
outbursts 
also in a (slightly)
less massive WD. The ejecta of RNe are expected to be less massive
than
those of CNe. This is because on the surface of the massive WD
expected in
a RN, the critical conditions for ignition are reached with a far 
less
massive envelope. Studies of the ejecta of RNe indicate a mass between
10$^{-6}$-10$^{-7}$ M$_{\odot}$, instead of the 10$^{-4}$-10$^{-5}$
M$_{\odot}$ observed in ``classical"
novae ( Gehrz et al. 1998, Hernanz and Jose 1998, Starrfield 1999).

RNe represent a convenient laboratory  to compare the  predictions
of the TNR theory with the observations. From the
observed $\dot{M}$ and the observed duration of the inter-outburst
interval, in
principle,  one
can obtain a
direct
estimate of the total mass accreted (M$_{accr}$) between two
successive
outbursts.
This
quantity can be  compared with both the (theoretical) critical
ignition
mass M$_{ign}$ and the
 mass of the ejected shell M$_{ej}$, which can be estimated by
spectroscopic
methods (and, in principle, by photometric methods  if   dP/P
can be accurately
measured, Livio 1991). A similar comparison cannot be
made in the case of CNe because the amount of mass accreted prior
to
outburst is
badly determined due to  their  long inter-outburst
interval.

Observations of RNe in quiescence  and outburst can also be used
to
determine
the
secular balance between the total accreted mass M$_{accr}$ and that
ejected
in the
explosive
phase M$_{ej}$, and therefore to investigate the possible role of
RNe as
progenitors of SN Ia. The ejecta of RNe have an almost ``solar"
chemical
composition and are not enriched in ``heavy" elements, such as
carbon,
oxygen and neon. This has been taken as an indication that the massive
white
dwarf in RNe is not eroded, will gain mass after each cycle of
accretion and ejection, and will eventually explode as a Type Ia
supernova
(Hachisu and  Kato, 2002).

In Gilmozzi and Selvelli (2007) (hereinafter Paper I), we 
studied  the
UV spectrum  of T Pyx in
detail and  our main
conclusion was that the spectral energy distribution (SED) is
dominated by
an accretion disk  in the 
UV+opt+IR ranges, with a  distribution that, after correction for
reddening E$_{B-V}$=0.25,  is described by a 
power law  F$_{\lambda}$ = 4.28 $\times$ 10$^{-6}$ $\lambda$$^{-2.33}$  erg
cm$^{-2}$ s$^{-1}$ \AA$^{-1}$, while the continuum in the UV range  
can  also be  
well represented by
a single  blackbody of T $\sim$ 34000 K. The observed UV
continuum
distribution
of T~Pyx has remained remarkably 
constant in both slope and intensity during  16 years of {\sl IUE}
observations.

In the present study, we  use this
basic result from the IUE data and other 
considerations,  to constrain
the
relevant parameters of the  system and to attempt to understand
the 
elusive nature of its outbursts.
The five recorded outbursts of  T~Pyx occurred
in 1890, 1902, 1920, 1944, and 1966, with a mean recurrence time of
19$\pm$5.3 yrs (WLTO).  All
outbursts were  similar in photometric behavior and
characterized by a decay time t$_3$ $\sim$ 90$^{\rm d}$,  a speed class
that is
substantially slower than in other RNe.
\noindent
We  divide the paper into several self-contained sections.

In Sect. 2, we discuss 
the problem of  distance, which we determine to be 3500$\pm$500
pc. In
Sect.
3, we analyze the mass function of the system. Using observational
data and
other
considerations, we obtain $1.25 < M_1 < 1.4$ M$_\odot$, $M_2 \sim
0.24$
M$_\odot$, or
$M_2 \sim 0.12$ M$_\odot$ (depending on the as yet uncertain
period), and $i
=25\pm5$ degrees.
In Sect. 4, we derive the absolute magnitude at minimum M$_{\rm
v}$=2.53$\pm$
0.23.
In Sect. 5, we calculate the bolometric disk luminosity $L_{\rm
disk} \sim
70$ L$_\odot$.

In Sect. 6 we use various methods to determine the
mass
accretion rate.
The value depends on $M_{\rm WD}$, and is $\dot{M} \sim 1.3\ \times
10^{-8}$
M$_\odot$ yr$^{-1}$ for
$M_{\rm WD}=1.33$ M$_\odot$. In Sect. 7, we derive the accretion
rate
before the 1966
outburst, and find that it is about twice the value measured with IUE.
In
Sect. 8, we compare
the accreted mass before the last outburst as determined
from the $\dot{M}$ derived in Sect. 7, $M_{\rm accr} \sim 4.5\ \times
10^{-7}$
M$_\odot$, with
theoretical calculations of the ignition mass. We find good
agreement
(including the
inter-outburst interval) for $M_{\rm WD} \sim 1.37$ M$_\odot$. In
Sect. 9,
we compare
the observations of T~Pyx with the theoretical models of Yaron et
al
(2005), finding  good agreement for all parameters
with the exception of $t_3$.

In Sect. 10, we reanalyze the historical data of the outbursts and
reach
the
inescapable conclusion that the mass ejected in the explosion is
$M_{\rm
ej} \sim 10^{-4}-10^{-5}$ M$_\odot$, about a factor 
100 higher than the accreted mass. This discrepancy is
discussed in
detail in Sect. 11, and leads to the conclusion that
the WD loses mass during the ourbursts. In Sect. 12, we revisit
the
parameters of the
extended nebula around T~Pyx, and demonstrate why the nebula  cannot be
used to
derive
information
about the mass ejected during outburst. Section 13 discusses the
mass
balance and infers  
that  T~Pyx is not destined to become a SN Ia as  often
postulated for recurrent novae. Far from
increasing its mass towards the Chandrasekhar limit, the WD in
T~Pyx is
being eroded.

In Sect. 14, we discuss the hypothesis that the WD in T~Pyx
is
undergoing steady nuclear burning
on its surface, and show that it is incompatible with both
historical and
UV observations. In Sect.
15, we report about XMM observations confirming our analysis by
directly
disproving the hypothesis. 

In Sect. 16, we discuss the recurrence time, and 
predict that
the
long awaited next outburst will occur around A.D. 2025.
Section 17
reports
our  conclusions.

\section{The distance }

An accurate knowledge of  distance is fundamental to the
determination
of the disk luminosity  L$_{disk}$ (and M$_v$),  the mass
accretion rate
$\dot{M}$, and
the mass of the
observed nebula surrounding the system M$_{neb}$ (not to be
confused with
the mass ejected during outburst, M$_{ej}$).

In this section, we revisit the problem of determining the distance to T
Pyx. We adopt a new accurate determination of the
reddening,
  E$_{B-V}$=0.25 $\pm$ 0.02, which is based on the depth of the $\lambda$
2175 \AA~
  interstellar absorption feature (see Paper I).

An estimate of the absolute magnitude at maximum  M$_{V}^{max}$ can
be
obtained from the optical decay time
$t_2$ or $t_3$ by  extending to a recurrent nova such as T Pyx, the
applicability of the Maximum Magnitude versus Rate of Decline (MMRD)
relations obtained using data of classical novae.  

 This is justified by the fact that it is clearly established that
the same
kind of physical process for the outburst is involved i.e. a TNR on
an
accreting
WD. The suggestion that recurrent novae in M31 are  fainter
than
predicted by the MMRD curve (Della Valle and Livio, 1998), applies
only to
the RNe subset that, unlike T Pyx, displays a  short $t_3$. The
long $t_3$ of T Pyx, the spectral behavior during outburst and the
quantity of mass ejected (see Sect. 10) indicate that the same
mechanism
is operating as in CNe.

The optical decay time of T~Pyx is well
determined
(Mayall 1967;  Chincarini
and
Rosino 1969; Duerbeck 1987; WLTO 1987; Warner 1995) with  $t_2$
$\sim$ 62$^d$ and $t_3$ $\sim$ 92$^d$.
From these values, we derived four different estimates 
for M$_{v}^{max}$  ($-$7.01, $-$6.75, $-$7.00, $-$6.64)
using 
the S-shaped MMRD relation of Della Valle and Livio (1995), and the
two linear
and the one S-shaped relations of Downes and Duerbeck (2000), respectively.

An additional estimate of the distance can be obtained by assuming
that T
Pyx, as a moderately fast nova, radiated at near-Eddington
luminosity at
maximum. This provides
\begin{equation}
L_{Edd}/L_{\odot} = 4.6 \times 10^4\;(M_1/M_{\odot}-0.26)
\end{equation}
For a massive white dwarf (M$_1$ $\sim$ 1.35 M$_{\odot}$), this
corresponds to M$_{bol}^{max}$ $\sim$ -6.98.
Close to the  epoch of maximum  light the observed color index  showed 
 oscillations about  an average value of  (B-V)=0.35    (Eggen et
al. 1967, 
Eggen, 1968), or, after correction for reddening,
 (B-V)$_o$ $\sim$ 0.10.
 Van den
Bergh and Younger (1987) found that  novae at maximum light have
(B-V)$_o$ = +0.23 $\pm$ 0.06, while Warner (1976) reported that the
intrinsic
color of novae close to maximum is  (B-V)$_o$ $\sim$ 0.0.  The
intermediate
value for T Pyx was 
(B-V)$_o$ $\sim$ 0.10,  which is  indicative
of a temperature  lower than 10,000 K and suggests that, close to
maximum light, 
the nova radiated mostly in the optical. 
Since  the bolometric correction is quite small in this case (BC 
$\sim
-0.20$),   it follows that  M$_{v}^{max} = -6.78$, which is quite close 
to the values obtained above from the four MMRD relations.  

The mean value of the  five estimates for  M$_{v}^{max}$ is
$-$6.83,  with a
median value  $-$6.79 and a formal standard deviation $\sigma$ =
0.16. 
  The five estimates are intermediate between the value ($-$6.4)
reported by Payne-Gaposchkin (1957), Catchpole
(1968),  and Warner(1995) and the single value  ($-$7.47) given by Duerbeck
(1981).

 Adopting  M$_{v}^{max}=-6.81$, $A_V=3.15 \times E(B-V)$ = 0.79 and
m$_{v}^{max}$
  $\sim$ 6.7 (Catchpole 1968, Warner 1995, Chincarini and Rosino
1969), one
  obtains a distance d = 10$^{(1+(m-M-A)/5)}$ $\sim$ 3500 pc.
 The  uncertainties  in the individual  parameters
 (0.16 for M$_{v}^{max}$,
 0.1 for m$_{v}^{max}$,  0.06 for A$_V$) correspond to 
an
 uncertainty of  about 330 pc in the distance, i.e. to a
relative uncertainty   
of the order of 10 percent.

A further  estimate of the distance can be  obtained by comparing
the
dereddened
flux of T Pyx at 1600 \AA~ with the accretion disk 
models by  Wade and Hubeny (1998). 
As mentioned in Paper I, the most appropriate descriptions of the shape
and depth of the lines (models $ee$ and $jj$ at low
inclination angles) correspond to high values of white-dwarf
mass and accretion
rate, although they  provide  a continuum that is  too steep. 
In any case,   these  high M$_1$, high $\dot{M}$ models, (we have 
extrapolated
$\dot{M}$
to values as high as log $\dot{M}$=-7.5)  clearly show that  the 
 ratio F$_{1600}^{model}$/F$_{1600}^{obs.}$ is about 1000. Applying a
scaling of a
factor of  $\sim$
(1000)$^{0.5}$ = 31.5  to the model  distance (100 pc) would
reconcile these
two
values. The derived  value of $\sim$ 3150 pc is in fair agreement with
the
assumed  distance of 3500 pc.

We recall that a  range of distances to T Pyx
can
be found in the literature. Catchpole (1969) derived a lower limit
of 1050
pc
from  the equivalent width of the CaII K line (EW $\sim$ 400 m\AA)
and the
  EW-distance relationship derived by Beals and Oke (1953) for
stars close to the Galactic plane. The EW measurement of Catchpole is
however quite
  uncertain because of possible contamination by emission lines and
the low
  quality of the spectra (the CaII K line falls close to the edge of
the
spectrum
and
  only two spectra were of sufficient signal in that region).

 Catchpole (1969), using  the two spectra of the highest
signal-to-noise-ratio found
that the K line of calcium corresponds to radial velocity of +20 km
s$^{-1}$.
According to Catchpole (1969), this velocity agrees well with that 
of  material travelling with  the Galactic rotation in the direction of  T
Pyx; since  
 this direction is, however,  close to a node of the rotation curve,  the
Beals and
Oke (1953) method cannot be used to determine the distance.    

 Warner (1976),  used the calibration given by Munch (1968) for
stars  close to a node  in the Galactic rotation curve, and indicated that 
a
star with a measurement of 
 0.4 \AA~  for the K line should be at least at a distance of 2 kpc.

 In a  study of the IS Ca II K line
towards
O- and
B-type stars in the Galactic disk, Hunter et al. (2006), presented a new
calibration
of the
the total column density of Ca II K versus distance. In a plot of
log
N(K) versus log d, the estimated distance of 3500 pc (log d =
3.545)
would correspond to log N (CaII) $\sim$ 13.0. The EW of the K line
measured by Catchpole (1969) corresponds to a column density of 4.9
$\times$
10$^{12}$
(log N=12.69), which would correspond to a distance of 2200 pc.
However, this value of column density is a lower limit,
because
the medium is not optically thin in the K line, and  a yet higher 
value of distance is expected. 

 By
applying  T Pyx theoretical predictions for light curves of  CVs
in
outburst, Kato (1990)
estimated a distance close to 4000 pc, by comparing  the
 observed  magnitude at outburst with  the theoretical
absolute magnitude prediction at outburst.

 In the following, we assume a distance of $d=3500 \pm 350$ pc. A
discussion
of the
critical role played by 
 distance in the interpretation of the
nature
of T Pyx is given in Sect. 11.1.

\section{ The system parameters}
We aim to determine both the disk luminosity and the mass accretion rate
of T
  Pyx.
This  requires 
prior knowledge of the system inclination angle  $i$   and  the
mass of
the
primary  M$_1$. In the disk geometry, the inclination angle is critical to
the  estimate of the disk luminosity, while  M$_1$ and R$_1$ (a
function of M$_1$) are key parameters in the correlation  between $\dot{M}$ 
and
L$_{disk}$.  For most CVs, the determination of these and  
other parameters (M$_2$, P$_{orb}$,
 2K) entering  the mass function:
\begin{equation}
\frac{(M_2\cdot sini)^3}{(M_1+M_2)^2}=1.037 \times 10^{-7}\cdot
K_1^3\cdot P
\label{eq:massfun}
\end{equation}
is a  difficult task, and accurate solutions have been
obtained only for a few eclipsing systems. 

For T Pyx,  the situation, ``prima facie", does not appear  
encouraging
because no system
parameter (not even the orbital period) have been accurately measured.
Therefore,
one has
to start from a  restricted  range of values  corresponding to the most
accurately  known
parameters in
the mass function equation, based
on theoretical or semi-empirical considerations, and derive the
corresponding  range of
allowed  solutions for the unknown relevant parameters.

\subsection{The orbital period(s) and K$_1$}

There have been several attempts to measure the orbital period of
T Pyx, although, disappointingly, they have  resulted in a wide range of
values  (see Paper I); this is a clear indication that there is no definite
photometric clock associated with the system. 

Patterson et al. (1998)  detected  a stable photometric wave  at
P$_h$=1$^h$.829  $\pm$ 0$^h$.002  (a value  close to that of  the
``most probable" 
photometric period P$_h$ =
1$^h$.828 of  Schaefer et al., 1992), although this interpretation is
inconsistent with   the presence of  another signal at 2$^h$.635. 

The only spectroscopic determination of the period is that  by 
Vogt et al.
(1990), who, in a preliminary  study based on a large
number
(101) of spectra of limited resolution  ($\sim$3 \AA),  reported a 
spectroscopic period  of P$_h$ $\sim$ 3$^h$.439,  without quoting
uncertainties. 
This
spectroscopic period is almost twice the photometric value determined
by
Szkody and Feinswog (1988) in the infrared.
In other CVs, this behavior is generally interpreted as evidence of
ellipsoidal variations in the companion, in a system observed  at a
high
inclination angle. This is not the case for T Pyx because, 
as convincingly shown in Paper I, the
companion does not contribute 
to the J light curve, unless the distance is less than 200 pc.

In
the
following, we consider   system solutions for both 
P$_h$ =
1$^h$.829 (photometric, Patterson et al, 1998)  
and
P$_h$=3$^h$.439 (spectroscopic,   Vogt et al. 1990).
The inclusion of the spectroscopic period could be criticized 
since it is
the  result of a preliminary analysis that has not been confirmed 
by
subsequent  studies. However, given the complex photometric period 
structure
and in
the absence of a definite physical interpretation  of the stable
signal
with P$_h$=1$^h$.829  (that could even arise from the rotation of a 
magnetic
white dwarf), we believe that 
the
spectroscopic
results should be considered, since, in any case, they have not been 
challenged by
similar
studies.
   
Vogt et al. (1990)  
found that the projected orbital velocity K$_1$ of the primary, 
derived
from an analysis of radial velocities of the emission lines, is
approximately 
24$\pm$5
kms$^{-1}$.
Taking into account the limited resolution and the large number of
spectra,
this value is probably an upper limit.  We assume  
that this  value of  K$_1$ is representative of  the primary
star radial
velocity (independently of the period being considered), although we are
aware of
the  
problems  encountered in associating 
  radial velocity  changes with the motion of the WD.

\subsection{The mass of the primary}

Both theoretical considerations and observational determinations
(although
limited) indicate that the WD masses in classical nova systems are
about
1.0
M$_{\odot}$, i.e.  higher  than inferred  from the standard field WD
mass
distribution  (Ritter et al., 1991).  Theoretical models of
recurrent novae
require an even 
more massive white dwarf (WLTO, Shara 1989, Livio 1994). Therefore,
in the
following,
we  allow the mass of the primary, M$_1$, to vary between  1.25 and 
1.4
M$_{\odot}$.

\subsection{The mass of the secondary }

 In the absence of any direct information about the  secondary (the
accretion
disk  is the dominant luminosity source from the UV to the
infrared,  Paper
I),  we can estimate its mass M$_2$  on the basis of  the commonly
accepted assumptions that: a) the secondary fills its Roche  lobe,
  b) its radius R$_2$ is determined by the Roche geometry, and c)
the
  secondary star obeys a mass-radius relation valid for low mass main
sequence
  stars.
In this  standard description, the 
radius of the secondary is well approximated by the equivalent
radius
R$_{L}$ of the
Roche lobe itself, i.e. R$_2$ $\sim$ R$_{L}$ and the spherical
volume of
the secondary is
assumed to  equal  the volume of its Roche lobe. The equivalent
volume
radii of
the Roche lobe are given in tabular form by several authors (Plavec
and
Kratochvil 1964; Kopal 1972; Eggleton 1983; Mochnacky 1984).
For our purposes,  it is  convenient however to use the
analytical approximation by Paczinsky (1971), which is  valid for
M$_2$/M$_1$
$\le$
0.8:
\begin{equation}
R_L/A = 0.457256(M_2/(M_1+M_2))^{1/3}
\label{eq:RL}
\end{equation}
where we have introduced a new value for the constant (instead of
0.46224) because it yields results that differ by less than 1\%
from those tabulated by Eggleton (1983 ) and Mochnacky (1984).

\noindent
By combining equation \ref{eq:RL} with Kepler's third law:
\begin{equation}
A=0.50557 M_{tot}^{1/3} P_{h}^{2/3}
\end{equation}
one obtains an approximate relation between P, M$_2$, and R$_2$
($\sim$
R$_{L}$):  
\begin{equation}
P_h = 8.997 R_2^{3/2} M_2^{-1/2}
\label{eq:PRM}
\end{equation}
\noindent
In the crude assumption M$_2$ $\sim$ R$_2$, this provides  an
approximate
relation: M$_2$ = 0.111 P$_h$
by which a rough estimate of the mass of the secondary can be
obtained if
the orbital period is known.
\noindent
In general, however, the R-M relation for the low MS stars is
described more accurately by a relation of the form:
\begin{equation}
R_2 \sim \beta M_2^\alpha
\label{eq:alpha}
\end{equation}
By combining eq.  \ref{eq:PRM}  and  eq  \ref{eq:alpha}, we 
derive a
more 
general relation between  M$_2$   and P$_h$:
\begin{equation}
M_2=P_h^{(2/(3\alpha-1))} \cdot 8.997^{(2/(1-3\alpha))} \cdot
\beta^{(3/(1-3\alpha))}
\label{eq:alphabeta}
\end{equation}
Values for the parameters  $\alpha$  and $\beta$  of  the low  
mass main-sequence stars
were derived from observational data  and theoretical considerations. In
particular,
Warner (1995) assumed that R$_2$=M$_2^{13/15}$ (which provides an
approximate  fit to
the data set by Webbink, 1990) and derived M=0.065$\times$P$^{1.25}$, while
Smith and Dhillon (1998) found a constitutive relation R$_2$=0.91
M$_2^{0.75}$
and
derived  M=0.038$\times$P$^{1.58}$.

Patterson et al. (2005) derived an empirical
mass-radius
sequence for CV secondaries based on masses and radii
measured primarily for  the superhumping CVs. They found that
R$_2$=0.62$\times$M$_2^{0.61}$
for P
shorter
than 2.3 hours, and R$_2$=0.92$\times$M$_2^{0.71}$ for P longer than 2.5
hours, and
derived  M$_2$=0.032$\times$P$^{2.38}$ (P $\le$
2.3)
and
M$_2$=0.026$\times$P$^{1.78}$ (P $\ge$ 2.5).
 We  inserted the $\alpha$ and
$\beta$ values of Patterson et al.(2005)  in Eq. \ref{eq:alphabeta}
and
obtained 
slightly
different results, that is, M$_2$=0.02823 P$^{2.41}$ (short P) and
M$_2$=0.02554 P$^{1.77}$ (long P).

\begin{table}
\caption{M$_2$ for P$^h$=1.829 and P$^h$=3.439  from  various
M$_2$-P
relations or   different $\alpha$ and
$\beta$ values in the literature (see text).}
\begin{center}
\renewcommand{\tabcolsep}{0.3cm}
\begin{tabular}{lcc}
\hline
 &P$_h$= 1.829  &    P$_h$=3.439  \\  
\hline                                                                            
Patterson et al.   & 0.134 &  0.234\\                                                                                
Smith and Dhillon  &  0.100 &  0.267 \\                                                                         
Knigge   &  0.140 &  0.236\\                                                                                           
Warner   &  0.138 &  0.304 \\                                                                        
Equation (7) &     0.121 &  0.227 \\

\hline
\end{tabular}
\end{center}
\end{table}

Knigge (2006) obtained an independent
M$_2$-R$_2$
relation by
revisiting and updating the mass-radius relationship for CV
secondaries
determined by Patterson et al. (2005). We have fitted with a
power-law the
data contained in his Table 3  and derived M$_2$-P
relations
for short P and long P systems. The relations are
\begin{eqnarray}
M_2 = 0.0371 \times P^{2.2022}~~~~~~~~~(P \le 2.2), \\  \nonumber 
M_2 = 0.02024 \times P^{1.9802}~~~~~~~~(P \ge 3.2).
\end{eqnarray}
Table 1 provides the M$_2$ values obtained by  these various methods,
for
P$_h$=1$^h$.829 (photometric) and
P$_h$=3$^h$.439 (spectroscopic).

In the following, we  therefore study the   system solutions for 
M$_2$ 
= 0.12 $\pm$0.02 M$_{\odot}$  and M$_2$ = 0.24 $\pm$ 0.02
M$_{\odot}$, 
separately. The corresponding spectral types are in  the range
M3V-M5V
(Kirkpatrick et al. 1991, Legget 1992).

\subsection{The  system inclination  }

The above derived range of allowed values for the system parameters
(M$_1$:
1.25-1.4 M$_{\odot}$, M$_2$ = 0.12 M$_{\odot}$ (for
P$_h$=1$^h$.829), 
M$_2$
= 0.24 
M$_{\odot}$ (for P$_h$=
3$^h$.439), K$_1$: 24
$\pm$5 km s$^{-1}$) can be inserted into Eq. (2)
to find the
corresponding range
of solutions for the system inclination, which is considered to be  a
free
parameter.

Table 2 clearly shows that, given the observational and theoretical
constrains for P, K$_1$, M$_1$, and M$_2$, the solutions for the
inclination
$i$ are  in the range between 20 and 30 degrees,
a value close to
25 degrees being the most likely value. In particular, for  M$_1$
$\sim
$1.35 M$_{\odot}$, (see also the following) the
solutions for $i$ do not change significantly, and for
P$_h$=1$^h$.829
(M$_2$=0.12 M$_{\odot}$) and P$_h$=3$^h$.439
 (M$_2$=0.24 M$_{\odot}$ ) are close to  30 and  20 degrees,
respectively (see also Fig. 1).

Patterson et al. (1998) suggested that T Pyx is
being observed at low system
inclination,  i.e.  $i$ $\sim$10$^o$-20$^o$, due to  the low
amplitude of the orbital signal. 
Shabbaz et al. (1997) estimated the binary inclination $i$ of T
Pyx by
measuring the separation of the H$_{\alpha}$ emission-line  peaks,
and
obtained a lower limit of $\sim$ 6$^o$.  
A low system inclination  is  also consistent with the small value of
the
radial velocity,
the sharpness of the emission lines in the optical (Warner, 1995),
and 
the steepness of the UV and optical continua (Paper I).  However, 
the
presence of radial velocity variations (Vogt et al. 1990), and the
modulations in the photometric variations preclude an $i$ value close
to zero.

\begin{table}
\caption{The system inclination  for M$_1$ = 1.25-1.40 M$_{\odot}$
and
K$_1$ =24 $\pm$ 5 km s$^{-1}$ for the cases (1) P$_h$=1$^h\!$.829,
M$_2$=0.12; and (2)
P$_h$=3$^h\!$.439,
M$_2$=0.24}
\begin{center}
\renewcommand{\tabcolsep}{0.2cm}
\begin{tabular}{cccc}

\hline
M$_1$  &    K$_1$  & $i$ (P=1$^h\!$.83)   & $i$ (P=3$^h\!$.44) \\  
(M$_{\odot}$)  &(km s$^{-1}$) & \multicolumn{2}{c}{(degrees)}\\
                                                                             
\hline                                                                            

1.25 & 19 & 22.9 & 14.7 \\
1.25 & 24 & 29.4 & 18.7\\
1.25 & 29 & 36.4 & 22.8\\
1.30 & 19 & 23.5 & 15.0\\
1.30 & 24 & 30.2 & 19.1\\
1.30 & 29 & 37.4 & 23.3\\
1.35 & 19 & 24.0 & 15.4\\
1.35 & 24 & 31.0 & 19.6\\
1.35 & 29 & 38.4 & 23.9\\
1.40 & 19 & 24.6 & 15.7\\
1.40 & 24 & 31.7 & 20.0\\
1.40 & 29 & 39.5 & 24.4\\

\hline
\end{tabular}
\end{center}
\end{table}

\begin{figure}
\centering
\resizebox{\hsize}{!}{\includegraphics[angle=-90]{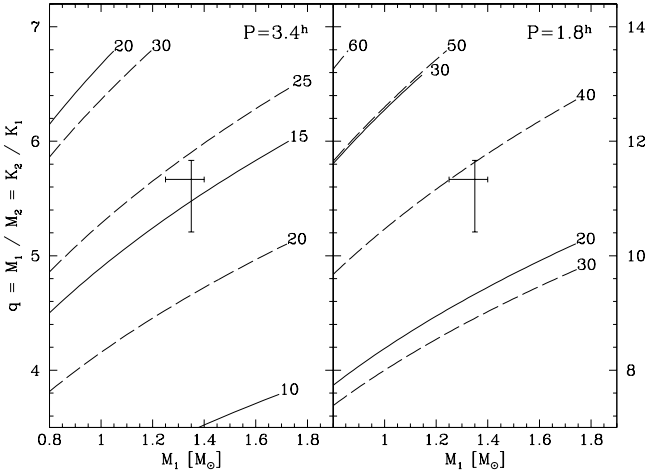}}
\caption{The $q \ vs \ M_1$ plane for P$_h$=3$^h$.439 and
P$_h$=1$^h$.829.
The lines of constant
inclination refer to $K_1=19$ km s$^{-1}$ (solid) and
$K_1=29$ km s$^{-1}$ (dashed), spanning the error
range in the value of $K_1$.  The values of $i$ are indicated.
The
error bars represent the ranges in $M_1$ and $M_2$
discussed in the text. }
\end{figure}

It should therefore be noted that, in spite of
the large
uncertainties in the observational data, the adoption of plausible
theoretical assumptions and  some semi-empirical 
constraints has enabled a quite restricted range for  the values of the
system
inclination to be defined. 
\noindent
 In the
following we will
assume $i=25\pm5
$ degrees for the system inclination angle.

\section{The absolute magnitude at minimum}

During the quiescent phase after
the
1967
outburst, the average optical magnitude of T Pyx   is
 m$_v$ = 15.3 $\pm$ 0.05 (Warner 1995, Duerbeck 1987, Schaefer
1992); for d =
3500 
$\pm$ 350 pc  and A$_V$=0.79 $\pm$ 0.06,
we obtain an (observed) absolute magnitude at minimum
M$_v^{obs}$=1.79 $\pm$ 0.21 (Warner 1995 defined this
to be the {\it apparent} absolute magnitude since the observed flux
in quiescence depends on the inclination angle).

Assuming that the visual radiation originates in a non-irradiated disk
(Paper
I), we correct the apparent absolute magnitude
M$_v^{obs}(i)$  for the
inclination
 angle of the disk (see Warner 1987, 1995) by a term  $\Delta{M_v(i)}$ to
obtain the ``standard" or ``reference" absolute magnitude: 
 \begin{equation}
 M_v^{corr}=M_v^{obs}(i)-\Delta M_v(i)
\end{equation} 
where $\Delta M_v(i)= - 2.5\log(2\cos i)$ according to WLTO (1987),
or 
 $\Delta M_v(i)= - 2.5\log[(1+1.5\cos i)\cos i]$ 
according to Paczynski and Schvarzenberg-Czerny (1980).

For $i\sim 25^o$, one  obtains a correction of $\Delta{M_v(i)}$
 $\sim$ 0.74 $\pm$ 0.06,
by averaging these two relations, and  $M_{v}^{corr}$=
+2.53 $\pm$
0.22 for the
absolute
magnitude,
averaged over all aspect angles.  This value is  about 1.4
magnitudes brighter than the mean absolute magnitude of nova
remnants in
the same
speed class (see Fig.
2.20 and Table 4.6 of Warner 1995). It is also  about 2.0
magnitudes brighter than the mean absolute magnitude of novae at
minimum,
as
obtained from the values given in Table 6 of Downes and Duerbeck
(2000).

The T Pyx remnant is therefore one of the brightest nova remnants, its 
absolute magnitude  M$_v^{corr}$=+2.53
being close to that of the ex-nova HR Del, 
whose corrected value, M$_v^{corr}$=2.30, implies
that it is a quite luminous object (see also Paper I).

\section{The disk luminosity from IUE and optical data}

As shown in Paper I, the integrated UV continuum flux of T Pyx in
the
wavelength range 1180-3230 \AA~, after correction for
reddening, is
1.94
$\times$ 10$^{-10}$ erg cm$^{-2}$s$^{-1}$.  For $d$ = 3500 $\pm$ 350 pc,
the
corresponding
integrated luminosity of the UV continuum is:
\begin{displaymath}
L_{UV} \sim 2.85\;\times\;10^{35}\: erg\; s^{-1} \sim 74.2\; \pm 15.0\;
L_{\odot}.
\end{displaymath}
 where the  20 percent is caused by  the 10 percent
uncertainty in the distance.
Most old novae have L$_{UV}$ in the range 1-20
L$_{\odot}$,
with V 841 Oph at $\sim24$ L$_{\odot}$ (Gilmozzi et al., 1994), and
V603
Aql 
and RR
Pic at $\sim11$ L$_{\odot}$  and $\sim3$ L$_{\odot}$, respectively 
(Selvelli, 2004).
Therefore, T Pyx is also quite bright  in the UV and, again, is
challenged
only by
HR Del that has L$_{UV}$ = 56 L$_{\odot}$ (Selvelli and Friedjung
(2002).

In the following,  we  assume that the observed UV and optical
luminosity
arise from an accretion disk heated by viscous dissipation of
gravitational energy. This  agrees with the 
behavior of
similar objects, like old novae and nova-like stars. This
assumption is
also strongly supported by the fact that the continuum energy
distribution
in the UV range is  well reproduced (Paper I) by a power-law
with spectral index $\lambda$  $^{-2.33}$, as predicted by
theoretical
optically
  thick accretion disk models radiating as a sum of  blackbodies.

The bolometric disk luminosity $L_{disk}$ can be estimated from the
observed UV and optical luminosity, where the bulk of the continuum
radiation is emitted, after correction for the inclination and 
the unseen luminosity in both the infrared and at
$\lambda$ $\le$ 1200 \AA. The  radiation emitted at wavelengths
shorter
than
Ly$_{\alpha}$ is
strongly absorbed and the energy is redistributed to longer
wavelengths
(Nofar, Shaviv and Wehrse 1992,  Wade
and  Hubeny  1998).

The integration (from $\lambda$ $\sim$ 1000 \AA~ to the IR) of the
power-law
distribution  which represents the observed UV and optical continuum of T
Pyx,
corresponds to a
flux of
about 3.6 $\times$ 10$^{-10}$ erg cm$^{-2}$ s$^{-1}$, which corresponds to
a
bolometric luminosity of 5.24 $\times$ 10$^{35}$ erg s$^{-1}$  $\sim$  
136.5
L$_{\odot}$.
This value refers to the adopted inclination of about 25 degrees.
After
correction to  the ``standard" inclination of about 57 degrees and by
considering  an
average limb-darkening factor similar to that in the optical, we
obtain an  angle (4$\pi$) averaged bolometric disk luminosity
(hereinafter
L$_{disk}$) of  about 70  $\pm$ 15 L$_{\odot}$, where the
relative
uncertainty (21 percent)  derives 
from the combination of the uncertainties in the distance and 
the 
inclination. 
We consider this to be  the reference disk luminosity for the
``recent",
post-1967-outburst epoch. 
 We note that the Wade and Hubeny (1998) models indicate  (their
Fig. 8) that in the
UV
region close to 1200 A (where the effect is larger)  the
limb-darkening  correction, from $i$ of about 25 degrees  to $i$ 
of about 60
degrees, 
corresponds to a
flux reduction of about 30 percent (this is limb darkening
alone, and the geometric projection factor  cos $i$ is not
included).  

A rough estimate of the 4$\pi$ averaged bolometric disk luminosity
can
  also be obtained  from M$_{v}^{corr}$ =  +2.53. Adopting an average disk
temperature 
of  about   28,000 K  (the UV continuum indicates  $\sim$ 34,000 K,
although this is too hot to match the optical fluxes), the corresponding
bolometric
correction
is
  $B.C.  \sim -$2.7,  and $L_{bol}$ $\sim$ 92.9  $L_{\odot}$,
in
fair agreement with the previous estimate. In this comparison, we 
neglected the effects of  temperature stratification in the accretion disk.

\section{The mass accretion rate}

One could in principle estimate the
  mass accretion rate $\dot{M}$ by comparing the observed spectral
  distribution with that of proper accretion disk models.  This
approach is
  not viable in our case for the following reasons: a) the spectral
index
of
  model disks that consist of  blackbodies is hardly sensitive to
the
mass
  accretion rate $\dot{M}$; b) most models providing $\dot{M}$ as a
function
  of the continuum slope do not include the case of a massive WD
with high
  accretion rates; c) disk models depend on a large number of
parameters,
so
  that model fitting does not generally provide unique results.

Alternatively,   if, as in our case, one can make the reasonable
assumption
  that the disk is heated by viscous dissipation
of
gravitational energy, the mass accretion rate $\dot{M}$ can be
obtained
from the relation:
\begin{equation}
\dot{M} = (2R_1 L_{disk})/(GM_1)
\end{equation}
\noindent

With
this method, $\dot{M}$ is not model dependent  but 
the knowledge of   $L_{disk}$ and  $M_1$  is required.

\noindent
Numerically, $\dot{M}$ can be represented by:
\begin{equation}
\dot{M}=5.23 \times 10^{-10} \phi  L_{disk}/L_{\odot}  \quad \rm{with}
\end{equation}
\begin{equation}
\phi=(R_1/M_1)/(R_{1o}/M_{1o}) = 0.1235 \cdot R_1/M_1,
\end{equation}
\noindent 
where R$_1$ is 
the
WD radius in 10$^{-3}$
R$_{\odot}$, R$_{1o}$ = 8.10 $\times$ 10$^{-3}$ R$_{\odot}$ is the radius
of a WD of mass  M$_1$ = 1.0 M$_{\odot}$, and  M$_1$ is in solar masses.

We  obtained average values for R$_1$  as a function of M$_1$  
from
various WD
radius-mass relations in the literature (Hamada and Salpeter  1961;
Nauenberg  1972;
Anderson 1988; Politano et al. 1990;   Livio 1994). 
By fitting a  quadratic function 
to these average values, for M$_1$ in the range  1.0 to 1.4
M$_{\odot}$, 
we found that:
\begin{equation}
R_1/R_{\odot}=  -0.01315M_1^2 + 0.01777M_1 + 0.00347.
\end{equation}
The upper curve in Fig. 2 is a plot of the standard R$_1$ $\sim$
M$_1^{-1/3}$ relation, which is generally valid for M$_1$ $\leq$ 1.0 
M$_{\odot}$,
while the lower curve  represents the quadratic fit
to the average values  for M$_1$  in the range  1.0 to 1.4
M$_{\odot}$,
 with a relative 
uncertainty of  $\sigma_{R_1}$/R$_1$ = 0.07.

\begin{figure}
\centering
\resizebox{\hsize}{!}{\includegraphics[angle=-90]{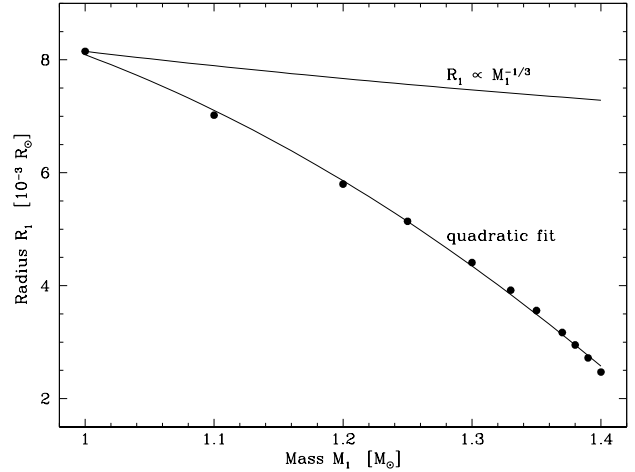}}
\caption{The lower curve represents  a quadratic
fit to the average values of  the WD radius  as a function of M$_1$ 
derived from various WD
radius-mass relations in the literature  (dots). The upper
curve is the
standard R$_1$ $\sim$ M$_1^{-1/3}$ relation.  }
\end{figure}

\noindent
Columns 2, 3, and 4  in Table 3 indicate  R$_1$,  $\phi$, and
$\dot{M}$ for $M_1$ in the range 1.00 - 1.4 $M_{\odot}$. We  note that
$\phi$ = 1.00 for M$_1$=1.0 M$_{\odot}$, and decreases to
about 0.23
for M$_1$=1.4 M$_{\odot}$.
The accretion rate $\dot{M}$ in Table 3 is calculated for the
(present) 
luminosity $L_{disk}$
$\sim$ 70 L$_{\odot}$, derived from  UV data. 
In the case of a massive WD, the mass accretion
rate is close to 1.1 $\pm$ 0.3 $\times$ 10$^{-8}$ M$_{\odot}$    $yr^{-1}$.
 The uncertainty in $\dot{M}$ is of the order of 23 percent
and derives from the 
uncertainties in the distance, the  correction for the
inclination, and
the value of R$_1$.    

\begin{table}
\caption{ M$_1$, R$_1$, the factor $\phi$
=(R$_1$/M$_1$)/(R$_{1o}$/M$_{1o}$),  the
post-1967 mass accretion
rate $\dot{M}$ (for L$_{disk}$ = 70 L$_{\odot}$), the ignition mass, 
and
the recurrence time $\tau$=M$_{ign}$/$\dot{M}$ (see Sect.
8). }
\begin{center}
\renewcommand{\tabcolsep}{0.1cm}
\begin{tabular}{cccccc}

\hline

M$_1$  &    R$_1$  &    $\phi$  &  $\dot{M}$  & M$_{ign}$  &  
$\tau$  \\  
(M$_{\odot}$)  & (10$^{-3}$R$_{\odot}$)& 
&(10$^{-8}$M$_{\odot}$yr$^{-1}$)
& (10$^{-7}$M$_{\odot}$) &  (yrs)\\ 
                                                                             
\hline 
\\                                                                           
1.00  &   8.10  & 1.000 &  3.66&  108.78  &  297.2   \\                                                                                
1.05  &   7.63  & 0.897 &  3.28&   92.88  &  283.2     \\                                                                         
1.10  &   7.10 &  0.797 &  2.92&   77.75  &  266.3   \\                                                                                           
1.15  &   6.51 &  0.699 &  2.56&  63.32   &  247.3    \\                                                                        
1.20 &    5.85 &  0.602 &  2.20&  49.19   &  223.6    \\                                                                          
1.25  &   5.14 &  0.508 &  1.86&  35.21   &  189.3    \\                                                                                  
1.30  &   4.35 &  0.413 &  1.51&  22.22   &  147.1   \\                                                                             
1.33  &   3.85 &  0.357 &  1.31&  14.78   & 112.8      \\                                                                       
1.35  &   3.51 &  0.321 &  1.17&  10.35   &  88.5       \\
1.36  &   3.33 &  0.302 &  1.11&  8.26   &  74.4    \\   
1.37  &   3.15 &  0.284 &  1.04&  6.35   &   61.1      \\                                                                           
1.38  &   2.96 &  0.265 &  0.97&  4.61   &  47.5     \\
1.39  &   2.78 &  0.247 &  0.90&  3.06   &  34.0     \\                                                                                 
1.40  &   2.60 &  0.229 &  0.84&  1.78   &  21.2    \\   

\hline
\end{tabular}
\end{center}
\end{table}

An independent confirmation of these values for the mass accretion
rate can
be obtained from the mass accretion rate versus boundary layer
luminosity  ($\dot{M}$-L$_{BL}$)  relation given in Table 2 of
Patterson
and Raymond (1985), if one assumes that L$_{BL}$ $\sim$
L$_{disk}$.
Their Table 2  indicates  that (for M$_1$=1.0)  a mass accretion rate 
$\dot{M}$
of 
1.6
$\times$ 10$^{-8}$
M$_{\odot}$ $yr^{-1}$, close to that derived above, would
correspond
to
a luminosity
of 1.3 $\times$ 10$^{35}$ erg s$^{-1}$. With the same $\dot{M}$, a  larger
luminosity is
expected by extrapolation to M$_1$ values close to 1.3 M$_{\odot}$,
in
close
agreement with the value found for L$_{disk}$ (2.70 $\times$ 10$^{35}$ erg
s$^{-1}$).

The L$_{1640}$-$\dot{M}$ relation reported in the same
Table 2
does not, however, provide 
a consistent result. The average (de-reddened) flux on earth in the
HeII
1640 line is 6.32 $\times$ 10$^{-13}$ erg cm$^{-2}$ s$^{-1}$ (Paper I) and
the
corresponding
luminosity for d=3500 pc is 9.3 $\times$ 10$^{32}$ erg s$^{-1}$. Table 2 of
Patterson and
Raymond  (1985) indicates instead a L$_{1640}$ = 1.2 $\times$ 10$^{32}$ erg
s$^{-1}$ for
$\dot{M}$=10$^{18}$ g s$^{-1}$=1.587 $\times$ 10$^{-8}$ M$_{\odot}$
yr$^{-1}$,
M$_1$=1,
and
extrapolation to higher M$_1$ values does not appear to be  able of 
reproducing 
the
observed HeII luminosity.  In T Pyx there is probably a 
nebular contribution to L$_{1640}$ that is added to that produced
by the
BL. It
should also be emphasized  that the HeII 1640 emission line exhibits
significant changes, while the UV continuum remains constant in
intensity and slope (see Paper I). This behavior casts  doubts
about the
suitability  of the L$_{1640}$-$\dot{M}$ method for the specific
case of T
Pyx.

Another estimate of $\dot{M}$ can be obtained by  direct
application of
the  M$_v$-$\dot{M}$ relations  reported in the literature.
\noindent
WLTO (1987) derived  a relation between the absolute optical
magnitude
and the accretion rate for the ``reference" disk model (Warner,
1987),
after
correction for the disk inclination: 
\begin{equation}
M_v^{corr.}=-9.48 -5/3 \cdot log(M_1\cdot\dot{M})  
\end{equation}
where   M$_1$
is in solar
masses and
$\dot{M}$ is in solar masses per year. In the case of T Pyx, 
for M$_1$ in the range 1.25-1.4 M$_{\odot}$,  one obtains
values of  $\dot{M}$ close to 5.0 $\times$ 10$^{-8}$ M$_{\odot}$ yr$^{-1}$.
It should be noted, however, that in the M$_v$-$\dot{M}$ relation
given by
WLTO there is no explicit dependence  on R$_1$, which strongly
depends on M$_1$, especially for massive WDs.
If $\dot{M}$ is corrected for the factor 
$\phi$=(R$_1$/M$_1$)/(R$_{1o}$/M$_{1o}$)
one obtains that $\dot{M}$(1.25) = 3.50
$\times$ 10$^{-8}$ M$_{\odot}$ yr$^{-1}$
and $\dot{M}$(1.40) = 1.60 $\times$ 10$^{-8}$  M$_{\odot}$ yr$^{-1}$. If
one
instead
adopts the
Paczynski and Schvarzenberg-Czerny (1980) correction for $i$,  
the corresponding
$\dot{M}$
values are 2.85 $\times$ 10$^{-8}$ M$_{\odot}$ yr$^{-1}$ and 1.22 $\times$
10$^{-8}$
M$_{\odot}$ yr$^{-1}$  respectively.

Lipkin et al. (2001) improved the $\dot{M}$-M$_v$ relations
presented by Retter and Leibowitz (1998) and Retter and Naylor
(2000) and derived the new relation 
\begin{equation}
\dot{M}_{17} = M_1^{-4/3} \times 10^{(5.69 - 0.4M_v^{corr})}  
\end{equation}

\noindent
where $\dot{M}_{17}$ is the mass transfer rate in 10$^{17}$ g
s$^{-1}$ and
M$_v^{corr}$ is
the
inclination-corrected absolute magnitude of the disk.
The factor M$_1^{-4/3}$ comes from the factor R$_1$/M$_1$ in the
assumption,
generally valid for WDs with M $\leq$ 1 M$_{\odot}$, that
MR$^3$=const.
However, this R$_1$-M$_1$ relation (polytropes) is not applicable if
the WD is
massive  because
in this case R$_1$ decreases with a far
steeper slope as M$_1$ increases  (see Sect. 6 and Fig. 2). 
The adoption of the standard R$_1$ $\sim$ M$_1^{-1/3}$ relation for
massive WD has the effect of overestimating R$_1$ and in turn 
overestimating
$\dot{M}$. This is especially true for M$_1$ $\sim$ 1.4, for which the
difference
between the R$_1$ values estimated from the R$_{1}$  $\sim$
M$_1^{-1/3}$
relation and  the more general  R$_{1}$ - M$_1$ relations
(Hamada and Salpeter  1961;
Nauenberg  1972;
Anderson 1988; Politano et al. 1990;   Livio 1994) reaches its maximum 
value (Fig. 2).

The Lipkin et al.  (2001) formula corrected for the proper
R$_1$/M$_1$
ratio
in
massive
white
dwarfs according to the $\phi$ factors of Table 3,  provides values of
$\dot{M}$ = 1.38 $\times$ 10$^{-8}$ and
0.59 $\times$ 10$^{-8}$ for white dwarfs with M$_1$= 1.25 M$_{\odot}$ and
M$_1$=1.40
M$_{\odot}$,
respectively.
These values are about one half of  the corresponding values obtained
from the
WLTO formula. It should be noted, however, that, in the derivation
of
Lipkin et al. (2001) formula there is the  assumption that
L$_{v}$/L$_{bol}$
$\sim$ 0.14. In the case of T Pyx, this L$_{v}$/L$_{bol}$ ratio is
lower by
a factor of 
about two and therefore the $\dot{M}$ derived using their formula
is
underestimated by a similar
amount.
In conclusion, the various methods provide quite consistent results
and we can
confidently assume that the  mass accretion rate derived from
``recent" UV
and optical data 
is 1.86 $\pm$
0.4
$\times$ 10$^{-8}$
M$_{\odot}$$yr^{-1}$ for M$_1$=1.25, and 0.84 $\pm$ 0.2 $\times$ 10$^{-8}$
M$_{\odot}$$yr^{-1}$ for
M$_1$=1.40
M$_{\odot}$.
For an
intermediate value of M$_1$=1.33 M$_{\odot}$ we obtain that
$\dot{M}$ $\sim$
1.3
$\times$ 10$^{-8}$
M$_{\odot}$ $yr^{-1}$.

 Table 4 summarizes the adopted values of the basic parameters 
and  their estimated  errors.

\begin{table}

\caption{The basic quantities}
\begin{center}
\renewcommand{\tabcolsep}{0.5cm}
\begin{tabular}{ll}
\hline
Quantity  &  value  \\ 
\hline
M$_v^{max}$  &  -6.81 $\pm$ 0.16 \\
m$_v^{max}$  & 6.7 $\pm$ 0.1 \\
A$_v$        &  0.79 $\pm$  0.06   \\
distance        & 3500 $\pm$ 350 pc \\
m$_v$         & 15.30 $\pm$ 0.05 \\
M$_v$         & 1.79 $\pm$ 0.21 \\
$i$           &  25 $\pm$ 5 degrees \\
$\Delta$m$_v$(i) & 0.74 $\pm$ 0.06 \\
M$_v^{corr}$  & 2.53 $\pm$ 0.23 \\
L$_{UV}$     & 74 $\pm$ 15 L$_{\odot}$ \\
L$_{disk}$   & 70 $\pm$ 15 L$_{\odot}$ \\
M$_1$ &   $\sim$ 1.36 M$_{\odot}$  \\
$\dot{M}$ &  1.1   $\pm$ 0.25  $\times$ 10$^{-8}$ M$_{\odot}$  yr$^{-1}$\\
$\dot{M}_{pre-OB}$ &  2.2  $\pm$ 0.5  $\times$ 10$^{-8}$ M$_{\odot}$ 
yr$^{-1}$\\

\hline
\end{tabular}
\end{center}
\end{table}
\noindent

\section{The pre-1966-outburst  mass accretion rate}

Schaefer (2005)  published a large collection of 
quiescent B-magnitudes for the recurrent novae T Pyx and U Sco
using
information
from archival plates, from the literature, and from his own
collection of
CCD magnitudes. The photometric data of T Pyx start in 1890 and
therefore cover all of  the known inter-outburst intervals.

According to these data, during  the inter-outburst phase in the years
1945-1966,  T Pyx was a factor of 2 brighter  in the B band than
during the
present quiescent  phase after the 1967 outburst. Since no
significant
changes in the (B-V)
  color index were found during both of these quiescence phases, one
can
safely
  deduce that in 1945-1966 T Pyx was a factor of 2 brighter also in
terms
of
  visual and bolometric flux.
Using the WLTO and the Lipkin et al. (2001) relations, one finds
that the mass accretion rate in epochs pre-1966-outburst,
hereinafter 
$\dot{M}_{pre-OB}$, was about twice the values of $\dot{M}$
for  post
1967,  obtained in the previous section. 

Therefore, we estimate that the mass accretion rate 
 $\dot{M}_{pre-OB}$ during the last
pre-outburst interval (1945-1966)  was between
1.68 $\pm$  0.4 $\times$ 10$^{-8}$ M$_{\odot}$ $yr^{-1}$ and  3.72 $\pm$ 
0.8
$\times$ 10$^{-8}$ M$_{\odot}$ $yr^{-1}$, for M$_1$=1.25 and  
M$_1$=1.40 respectively.

{\bf These are the 
values that are compared in Sects. 8 and 9  with the
theoretical
models of nova.}
 The post-1967 accretion values  (Table 3)
are considered in Sect. 16 in the context of the ``missing" 
outburst and the next expected outburst.

\section{The theoretical ignition mass and the accreted mass}

A nova outburst occurs when, due to the gradual accumulation  of
H-rich
material on the surface of the white dwarf, the pressure  
at the bottom of
the accreted layer becomes sufficiently high  for nuclear ignition
of H to
begin ( Shara 1981, Fujimoto 1982, MacDonald
1983). Since the radius R$_1$ of the white dwarf varies
approximately as M$_1^{-1/3}$ for M$_1$ $\le$ 1.0 M$_{\odot}$, the
critical
pressure for ignition
\label{eq:PIGN}
\begin{equation}
P_{ign} = (G\cdot M_1\cdot M_{ign})/(4\pi R_1^4)
\end{equation}
\noindent
corresponds 
to a critical ignition  mass
M$_{ign}$
that decreases approximately as M$_1^{-7/3}$,
while,
for more massive WDs M$_{ign}$, decreases with a 
steeper 
slope (see Table 3 and Fig. 2). In any case,  massive white dwarfs
need to accrete a
small amount of mass to reach the critical conditions.

For a given M$_1$ value, the critical ignition mass has been
calculated by
various authors  and the
reported values (for a given M$_1$) differ from each other mostly
in
the
choice of the critical pressure at the base of the accreted
envelope, which
varies  between   2.0 $\times$ 10$^{19}$  and 6.0 $\times$ 10$^{19}$
dynes cm$^{-2}$ (see Gehrz et al. 1998; Starrfield et al. 1998;  
Livio and Truran 1992; Hernanz and Jose' 1998; Truran 1998). These
studies indicate  that a lower limit to $M_{ign}$ (for a massive white
dwarf
with
$M_1$ close to 1.4 $M_{\odot}$) is in the range  2.0 - 4.0 $\times$
10$^{-6}$
$M_{\odot}$. However, as first pointed out by MacDonald (1983) and
in 
more
detail by
 Shara (1989) and  Prialnik and Kovetz (1995), the behavior
of a CN
eruption
and in particular  the critical mass depends (apart from the WD mass) also
on the
mass accretion rate (since the WD is heated by accretion) and the
temperature of the isothermal white dwarf core.

Townsley and Bildsten (2004) confirmed  that the 
earlier
prescriptions for ignition, based on the simple scaling M$_{ign}$
$\propto$
R$^4$ 
M$_1^{-1}$  for a  unique P$_{ign}$, are inadequate
and that  a
system of a given M$_1$ mass can have value of M$_{ign}$ that varies by a
factor of 10 for
different $\dot{M}$. 
At the high $\dot{M}$
values typical of most CVs,
the critical pressure can decrease to values as low as 3 $\times$ 10$^{18}$
dyn
cm$^{-2}$,  and the critical mass decreases accordingly. Thus, 
 in a  massive white dwarf accreting at significantly high rates,   one
expects a
value
for  M$_{ign}$ as low as  2-4 $\times$ 10$^{-7}$ M$_{\odot}$.

The critical envelope mass $M_{ign}$  as a function of M$_1$
and
$\dot{M}$ can be numerically approximated
 to be  :

\begin{eqnarray}
log M_{ign}=-2.862+1.542\cdot M_1^{-1.436}ln(1.429-M_1)+\nonumber\\
-0.19(log\dot{M}+10)^{1.484}
\end{eqnarray}
(Kahabka and van Den Heuvel 2006) where M$_1$ is in M$_{\odot}$ and
$\dot{M}$ is in M$_{\odot}$  $yr^{-1}$.

Table 5 indicates (for various M$_1$ and hence R$_1$ values)
 $\dot{M}_{pre-OB}$, the theoretical $M_{ign}$, the accreted mass
M$_{accr}$=$\Delta$t$\cdot$  $\dot{M}_{pre-OB}$, where $\Delta$t=22
yrs is
the
pre-1966 inter-outburst interval,   and 
$\tau$=$M_{ign}$/$\dot{M}_{pre-OB}$, that is, the expected
recurrence time
in
years. The mass accretion rate  and  $M_{ign}$ were
calculated using the quadratic fit for  R$_1$ as a function of
M$_1$
for 
massive white dwarfs   derived in Sect. 6  (see also Table 3), and
not
the approximation R$_1$ $\propto$ M$_1
^{-1/3}$.   
Table 5 clearly shows that, after allowances for errors in the
estimate of
$\dot{M}_{pre-OB}$, 
the expected recurrence time $\tau$ = M$_{ign}$/$\dot{M}_{pre-OB}$ is
close
to
the
observed value (22 yr)  for M$_1$ $\sim$
1.36 - 1.38 M$_{\odot}$ corresponding to  M$_{ign}$ and  M$_{accr}$   
in the range 3.0 to  6.0   $\times$ 10$^{-7}$   $M_{\odot}$.

This agrees with the estimate of Kato (1990)  that the white dwarf mass
for T Pyx
is between 1.3 and 1.4 M$_{\odot}$, while in a subsequent paper
Kato
and Hachisu (1991)
assumed M$_1$=1.33 M$_{\odot}$.

We confirmed  the results for M$_{ign}$ using the approximate
relation
for the ignition mass as a function of M$_1$ and $\dot{M}$ given by
Kolb
et al. (2001):
\begin{equation}
M_{ign}=4.4 \times 10^{-4}\cdot R_1^4\cdot
M_1^{-1}\cdot\dot{M}^{-1/3}  
\end{equation}
(where M$_{ign}$ and M$_1$ are in M$_{\odot}$, R$_1$ is in 10$^9$
cm, and
$\dot{M}$ is in 10$^{-9}$ M$_{\odot}$ yr$^{-1}$) and obtained
M$_{ign}$ = 2.72 $\times$ 10$^{-7}$ M$_{\odot}$.
A similar result is also  obtained by graphical interpolation in
Fig. 5
of 
Kahabka and Van Den Heuvel (1997), which for M$_1$  $\sim$ 1.37
and  $\dot{M}$ $\sim$ 2 $\times$ 10$^{-8}$ M$_{\odot}$ yr$^{-1}$  provides 
an
ignition mass close to  3.0 $\times$ 10$^{-7}$ M$_{\odot}$ (we note,
however, 
that these parameters 
 fall within a  region in which  only weak flashes are expected). 
In addition, we  derived  P$_{ign}$ as a function of $\dot{M}$  
and
 the WD temperature from   Fig. 3 of Yaron et al. (2005), and
found  
that   P$_{ign}$ 
$\sim$  4.8 $\times$ 10$^{18}$ for $\dot{M}$ = 2 $\times$
10$^{-8}$M$_{\odot}$
yr$^{-1}$  
(we note that in this regime there is a  weak
dependence on the WD temperature).
The insertion of this value of  P$_{ign}$  in equation
\ref{eq:PIGN}
gives  M$_{ign}$ = 3.8 $\times$ 10$^{-7}$ M$_{\odot}$.

The consistency of all these results  confirms 
that in the case of T Pyx the  ignition mass was close to 4.5
$\times$ 10$^{-7}$
M$_{\odot}$.

\begin{table}
\caption{M$_1$, the estimated pre-1967-outburst accretion rate
$\dot{M}_{pre-OB}$
(for
L$_{disk}$=140 L$_{\odot}$), the
theoretical ignition mass M$_{ign}$, the accreted mass
M$_{accr}$=22$\cdot$  $\dot{M}_{pre-OB}$ and the expected 
recurrence time
$\tau$=M$_{ign}$/$\dot{M}_{pre-OB}$. }
\begin{center}
\renewcommand{\tabcolsep}{0.2cm}
\begin{tabular}{ccccr}
\hline
M$_1$  & $\dot{M}_{pre-OB}$  & M$_{ign}$   & M$_{accr}$ & $\tau$ \\                               
(M$_{\odot}$)   & (10$^{-8}$M$_{\odot}$yr$^{-1}$)&
(10$^{-7}$M$_{\odot}$)& 
(10$^{-7}$M$_{\odot}$)    &
(yrs) \\ 
\hline
1.00  &    7.32  &   77.2 & 16.10    &105.5     \\
1.05  &    6.56  &   67.6 & 14.43    &103.0    \\
1.10  &    5.84  &   56.2 & 12.84   &96.2     \\
1.15  &    5.12  &   45.7 & 11.26   &89.2     \\
1.20  &    4.40  &   36.3 & 9.68   &82.5    \\
1.25  &    3.72  &   25.7 & 8.18    &69.1     \\
1.30  &    3.02  &   16.2 & 6.64    &53.6    \\
1.33  &    2.62  &   11.0 & 5.76   &42.0    \\
1.35  &    2.34  &   7.69 & 5.15    &32.8     \\
1.36  &    2.22  &   6.14 & 4.88     &27.7     \\ 
1.37  &    2.08  &   4.73 & 4.58    & 22.7      \\  
1.38  &    1.94  &   3.44 & 4.27    &17.7     \\
1.39  &    1.80  &   2.29 & 3.96    &12.7   \\ 
1.40  &    1.68  &   1.33 & 3.69    &7.9   \\   
\hline
\end{tabular}
\end{center}
\end{table}
\noindent

In the recurrent nova T Pyx, the theoretical ignition
mass
and  observed  accreted mass  
are in excellent agreement for a
massive WD, which  provides  new,  independent support  of the TNR theory.
Theoretical  expectations   have received  only   limited 
confirmation because  studies of the system parameters (in U
Sco,
T Cr B) have provided contradictory results.

\section{Comparison with the nova models of Yaron et al. (2005) }

The realization that three basic and independent parameters, the
 white dwarf mass M$_1$, the temperature of its isothermal core
T$_c$,
and
the mass transfer rate $\dot{M}$, control the behavior of a CN
eruption, and the improvements in computer power and  codes, has enabled
an
increase in sophistication in  simulating a nova outburst.
Prialnik
and Kovetz (1995) presented an extended grid of multicycle nova
evolution
models that have been extensively used by researchers. Each
observed nova
characteristic (e.g.  peak luminosity,  recurrence time, 
duration of the high luminosity phase,  outburst amplitude, 
mass of ejecta,  average outflow velocity, etc.) can be reproduced by 
a
particular combinations of  values of M$_1$,T$_c$, and
$\dot{M}$. Following this  earlier study, Yaron et al. (2005) extended
and
refined the resolution in the grid of models, including a
considerable
number of new parameter combinations. The full grid covers the
entire
range of observed characteristics, even those of peculiar
objects.

By matching the observed characteristics of a particular nova with
its theoretical counterpart, it is therefore possible to derive information
about the
mass and temperature of the white dwarf and its average accretion
rate.
Therefore, the grids in Tables 2 and  3 of Yaron et al.
(2005) can
be used to determine a set of fundamental parameters that
provides the
closest agreement with the observed data.

In the case of T Pyx, a comparison of the observed parameters for
the
eruption phase (the recurrence time $\tau$ , the outburst amplitude  A, the
decay rate t$_3$, the expansion velocity v$_{exp}$) with the
corresponding values
in the grid clearly indicates that the closest agreement corresponds
to values
 that are intermediate in between
1.25 and 1.40 for M$_1$, and -8.0 and -7.0 for log$\dot{M}$; at
these high
$\dot{M}$ values,
the effects of the core temperature T$_c$ are minor.
 Since a  denser grid is not
 available, we 
calculated  intermediate values by graphical and non-linear
interpolation in the published values. A satisfactory agreement 
with the
observed data comes from a model (model A) with M$_1$=1.36 and
log$\dot{M}$=7.5
($\dot{M}$=3.2 $\times$ 10$^{-8}$), which provides:
\noindent
$\tau$ $\sim$ 20, v$_{exp}$ $\sim$ -1200, A $\sim$ 7.6, t$_3$
$\sim$ 13, and 
$M_{ign}$  $\sim$ 5.0 $\times$ 10$^{-7}$.
\noindent
Following  our  interpolation of the  data, Yaron (2007)  kindly
provided us
with
 results obtained by  execution of the code for other
intermediate
models, that is, WD masses between 1.30 and 1.33, log$\dot{M}$
between -7
and
-8, and different WD temperatures. The closest agreement with the
observed
values originates in a model with M$_1$=1.33, T$_1$= 10$^7$,
log$\dot{M}$=-7.3
(model B), although
 the model with M$_1$=1.30, T$_1$= 10$^7$,  and
log$\dot{M}$=-7.15
(model C) is also acceptable.
In Table 6, we also include, as an illustration, a model
directly
obtained from  
the published Tables 2 and 3, for M$_1$= 1.4 M$_{\odot}$ 
and $\dot{M}$ = 1.0 $\times$ 10$^{-8}$ (model D). 

The value for $M_{ign}$ (about 7 $\times$ 10$^{-7}$
M$_{\odot}$) that
corresponds to the model of  Yaron et al. (2005) with M$_1$=1.33 and
log $\dot{M}=7.5$,
 is close to that
($\sim$ 6.0 $\times$ 10$^{-7}$ M$_{\odot}$) estimated by the observed
average
$\dot{M}_{pre-OB}$ (about
3.0 $\times$ 10$^{-8}$ M$_{\odot}$   yr$^{-1}$)
and the observed inter-outburst recurrence time of 22 yrs.

\begin{table}
\caption{A comparison between Yaron et al. (2005) grids  and 
observations.}
\begin{center}
\renewcommand{\tabcolsep}{0.1cm}
\begin{tabular}{lccccc}
\hline  
Param.  &  Model A&       Model-B &   Model-C &  Model-D  &  Obs.\\
\hline    

M$_1$  &    1.36         &   1.33   &      1.30   &     1.40   &     
1.35$\pm$ 0.05\\
$\dot{M}$ &  3.0  10$^{-8}$       &  5.0  10$^{-8}$& 7.0
 10$^{-8}$
&1.0
 10$^{-8}$ & 2.4$\pm$0.6  10$^{-8}$\\
M$_{ign}$ &   6.6  10$^{-7}$ & 1.03  10$^{-6}$ & 1.40
 10$^{-6}$&  2.0
 10$^{-7}$  &  $\sim $5.3  10$^{-7}$ \\
M$_{ej}$ &   6.4  10$^{-7}$       &   0.93  10$^{-6}$ &
1.30
 10$^{-6}$ &  
2.0  10$^{-7}$   &         (see Sect. 10.2)\\
$\tau$ &     20        &    20.6  &           19.8 &        
20.2 &        
22\\
Ampl. &        7.6         &  6.9  &            6.6 &           8.4
&  
8.0\\
v$_{exp,max}$ & -1500         &   -720  &           -572 &       
-1760  &  
-2000\\
t$_{3,vis}$ &    12        &      32   &           45   &        
$\sim$ 6
&  90\\
\hline
\end{tabular}
\end{center}
\end{table}

For T Pyx, the values of two of the
independent, theoretical parameters (M$_1$ and  $\dot{M}$) have
been
already constrained to within a limited range by theoretical
considerations 
(M$_1$)  or inferred by direct observations
($\dot{M}$). Since  the model grids
require similar values for M$_1$ and $\dot{M}$,  the results appear
to be consistent.
An additional  fine tuning of the model data could certainly
improve the fit but the overall agreement between models and
observations can be considered satisfactory:  an important 
parameter such as the recurrence time is reproduced quite well and for
the
other basic quantities the discrepancy is not large.

A parameter that deserves  special comment is   t$_3$. 
It  is not explicitly listed in the 
theoretical grid  but  can be associated to  t$_{ml}$, and was
included
by Yaron (2007) in  specific calculations. The observed
t$_3$  value
is about  three times longer than the expected  value.
Models with such
long t$_3$ would correspond to  higher M$_{ign}$ and therefore
to  far
more
massive ejecta. Models with a longer  t$_3$ (i.e. Model-C in Table
6) would be of too low v$_{exp,max}$ and  too high M$_{ign}$. 

In Sect. 10.2 we demonstrate that the spectroscopic observations
acquired 
during the 
last outburst in 1966   indicate the
ejection
 of a more massive shell
than that
expected from the simple  estimate based on  the quiescence
$\dot{M}_{pre-OB}$ and a
times interval  $\Delta$t  of 22 years.

Contini and Prialnik (1997) presented
evolutionary
calculations of TNRs on the surface of an accreting WD that
simulates
the observed outburst characteristics of   T Pyx and  found that
the model
that reproduces the observed characteristics of T Pyx most accurately is
obtained for an
accretion rate of 10$^{-7}$
M$_{\odot}$  $yr^{-1}$ onto a 1.25 M$_{\odot}$ WD;  the (theoretical)
mass
ejected in a single outburst
is expected to be close to  1.6 $\times$ 10$^{-6}$ M$_{\odot}$.   They 
also 
modeled the
nebular spectrum  and found that the theoretical fluxes agree with
those
derived from observations (Williams 1982) if  a distance of $\sim$
2 kpc 
is
assumed. However,  our detailed study of  the T Pyx 
luminosity and mass accretion rate  suggested  an
$\dot{M}_{pre-OB}$
value close to 3.7
$\times$ 10$^{-8}$ M$_{\odot}$  $yr^{-1}$, onto a 1.25 M$_{\odot}$ WD for
d $\sim$
3.5 kpc. If d $\sim$ 2.0 kpc, this would imply that  $\dot{M}$ $\sim$ 1.2
$\times$ 10$^{-8}$, a
factor of  10 lower
than  assumed by Contini and  Prialnik  (1997).
We note that a distance as large as about 7.0 kpc is required 
to derive  a luminosity correspomding to  $\dot{M}$ $\sim$ 10$^{-7}$.
See Sect. 11.1 for additional considerations of the distance.

\section{The photometric and spectroscopic history of the  outbursts}
\subsection{ The  long lasting optically thick
phase}
 
The five recorded outbursts of  T Pyx occurred in 1890,
1902, 1920, 1944, and 1966, with a mean recurrence time of
19$\pm$5.3 yrs
(see
WLTO, 1987).
A common feature of these outbursts was the far  longer  optical
decline time (t$_3$ = 90$^d$)  compared with that of other
 recurrent novae, which, with the exception of CI Aql (t$_3$ = 33 d)
and IM
Nor (t$_3$ = 50 d),
are generally much faster,  with t$_3$ of the order of  days.

 The well studied outburst of Dec. 1966 exhibited  a  sharp
initial rise to a shoulder (pre-maximum halt on Dec. 10, 1966) in  V
of about
7.9,  a near
flat maximum close to  V = 7.5 (which lasted about 30 days), a sharp
peak at V=6.8  on  Jan. 9, 1967, and  a slow decline with
t$_3$
$\sim$
90$^d$.
The first spectroscopic observations were obtained by Catchpole
(1969),
12.6 days after the initial halt. The spectrum was characterized by 
sharp
P Cyg features in the hydrogen lines and by other weak
emission lines. The presence of the absorption features endured
until  March 5, 1967,  and
since then they  became quite ``diffuse". Catchpole (1969), apart from
the
value of 850 kms$^{-1}$ at some epochs, reported outflow velocities
of the
order of
-2000 kms$^{-1}$ for both absorption and  emission
(permitted and
forbidden) components.

There
  is an observational gap of about 40 days  in
the spectra obtained by Catchpole, which was fortunately covered in
part
(from 09 Jan. 1967 to 16 Feb. 1967) by
the  spectroscopic data by Chincarini and Rosino (1969). They
described the
absorption system in the Balmer lines (up to H$_{12}$) as being
particularly
sharp and strong and noted an increase in the expansion velocity
from
-1535 km s$^{-1}$ on Jan. 31, 1967, to
-1760
km s$^{-1}$ on Febr. 2,  1967 and -1820 km s$^{-1}$ on Febr. 6 1967.
According to
Chincarini and Rosino (1969), the mean radial velocity during  the
period 
Jan.31-Feb. 6 determined from all  measurable absorption lines
was
-1810 $\pm$ 40 km s$^{-1}$.

Outflow
velocities of the order of 1500 - 2200 kms$^{-1}$ were also observed
in
previous outbursts.
Adams and Joy (1920)  reported the presence of ``dark" components
of
radial velocity
up to -2100 km s$^{-1}$ a
few days after maximum.  Joy (1945)  observed expansion
velocities of 1700 kms$^{-1}$ in forbidden emission lines in
spectra taken
some
months after outburst. Similar values were also found  by Herbig
(1945) and
reported by Payne-Gaposchkin (1957).

The ``principal absorption
system" is generally associated with the bulk of the mass ejected
during 
outburst, and for most novae has a velocity close to that
of
the ``nebular system" as determined from the width of the nebular
emission
lines  (Payne-Gaposchkin 1957, Mc Laughlin 1956,   Pottasch 
1959). The  v$_{exp}$ value reported by Joy (1945) referred
to observations during the  nebular phase, three-four months after
outburst. That the emission lines in the  nebular spectrum  showed
velocities of
the same order as
those deduced from the absorption lines  of the principal spectrum  
is an indication of near constant expansion.  

We note that in the literature on T Pyx little attention has been
paid  to
the fact that for almost three months after the initial halt T Pyx
showed a
strong continuum with the presence of emission and absorption 
lines  of similar strength.  The
persistence for almost three months  (about t$_3$) of  displaced
absorption 
components  in the H (and FeII) lines, which are observable by
combining the
spectroscopic observations of
Catchpole   (1969) and Chincarini and Rosino (1969) for the 1966
outburst, 
indicates an
 optically thick phase of similar duration.

In this respect, we recall that, before the outburst of
1966-1967,   Mc Laughlin (1965)  noted  that T Pyx 
was  exceptional among RNe since its photometric and spectroscopic
behavior
closely resembled that of a typical
nova both close to maximum (where it remained for several weeks to within
about a
magnitude) and in the nebular stage.

\subsection{The mass of the shell ejected in the optically thick
phase}

The similarity  between  the spectroscopic and photometric
characteristics of the 
outbursts of   T Pyx  and  those  of CNe, which allegedly eject
about 
10$^{-4}$ - 10$^{-5}$ M$_{\odot}$, suggests in itself that during
outburst
T
Pyx expelled a  shell of comparable  mass.

Classical novae undergo an optically thick phase  during
which they resemble each other, a fact that can be explained by the same
mechanism
(i.e. flux redistribution) producing the spectrophotometric
light
curve (Shore 1998, 2008). To achieve flux redistribution, 
the
material must  reach column densities of the order of 
10$^{23}$-10$^{24}$
cm$^{-2}$, 
which corresponds to  masses of about  
10$^{-4}$ - 10$^{-5}$
M$_{\odot}$.
  
An  optically thick stage also characterized  the outbursts of T
Pyx,  as
can be 
directly inferred from the lengthy period of time during which the optical
magnitude was close to its 
maximum value, with t$_3$ $\sim$ 90$^d$,  and from the presence of
absorption
lines of
HI and FeII, which lasted for at least 80 days.  

From the duration of the optically thick phase (associated to
t$_2$)  and the observed v$_{exp}$,  using simple assumptions, we
can
estimate the mass of the  shell ejected 
during outburst. The outer radius of the  shell can  be
estimated from the
observed 
expansion
velocity (v$_{exp}$ $\sim$1500
km s$^{-1}$)  and
the time elapsed from outburst, assuming continuous ejection. This
assumption 
 is justified by the persistence of displaced absorption components 
with similar equivalent widths. 
The  shell radius  is:
\begin{displaymath}
R_{ej} \sim
7.7 \times 10^{14} cm  \sim 1.1 \times 10^{4} R_{\odot}.
\end{displaymath} 
\noindent
The corresponding shell volume
is
V$\sim$1.84
$\times$ 10$^{45}$ cm$^3$.
\noindent
These values  are
conservative  because the terminal velocity was probably higher.
\noindent
To reproduce the optically thick stage  recognized 
from the presence and persistence of the absorption lines, the
column
density must be 
of the order of 10$^{23}$ cm$^{-2}$. Therefore, the
average density in the shell
must be close to 10$^{8}$ - 10$^{9}$ cm$^{-3}$, 
(we
note that  the density  should scale as R$^{-3}$). This
density
value agrees
with the spectroscopic behavior and the  presence of  permitted
emission
lines only. If we  assume that the ejecta are homogeneous and 
consist of
ionized  hydrogen,
the
mass  might be estimated as 
\begin{equation}
 M_H = N_e m_H V \sim 3.1 \times 10^{29} g
= 1.5 \times 
10^{-4} M_{\odot}.
\end{equation} 
This value is probably an upper limit because of
the
assumptions of continuous ejection and homogeneous shell. 

In an alternative approach, following  Williams (1994), one can
estimate
the hydrogen column density 
produced by an expanding
shell of  mass  1.0 $\times$ 10$^{-4}$ M$_{\odot}$ :
\begin{equation}
N_H\cdot R = 3.0 \times 10^{52}\cdot R^{-2}\quad [cm^{-2}]. 
\end{equation}
For T Pyx at
day 60 (= t$_2)$, we find 
that R$^2$=
5.8 $\times$ 10$^{29}$  [cm$^2$],
and N$_H$  $\cdot$R $\sim$ 5.2 $\times$ 10$^{22}$ [cm$^{-2}$]. A  mass of
the
ejecta higher
than
10$^{-4}$  is therefore required to produce an optically thick
stage until
day
60.

It is well established that in novae, the mass of the ejected
envelope  is
directly correlated with the optical decay time t$_2$ or t$_3$
(Livio,
1994). Therefore, 
an independent   estimate of the ejected  shell mass can be 
derived
from the relation: 
\begin{equation} 
log M_{ej}= 0.274 \pm 0.197 \cdot log t_2 - 4.355 \pm 0.283
\label{eq:dellavalle}
\end{equation}
\noindent
(Della Valle et al.
2002). Even after considering the   large uncertainties in
this
relationship,  a t$_2$=62$^d$ implies  a mass for the envelope 
M$_{ej}$
$\sim$
10$^{-4}$ M$_\odot$,  which is similar to the mass ejected by
classical
novae.

The data leading to Eq.  \ref{eq:dellavalle}  suffer from a large
scatter. We suspect that  the ejecta  expansion velocity  plays
also an
important role and should be included in the relation. 
Shore (2002, 2008)  suggested an approximate scaling
relation
for the 
optically thick stage:
\begin{equation}
M_{ej} \sim  6.0\ \times 10^{-7} \ \epsilon \ N_{H,24} \ V_3^2 \ t_3^2 \
M_{\odot},
\end{equation}
where V$_3$ is the outflow velocity in 10$^3$ km s$^{-1}$, and 
$\epsilon$
is the filling factor that, for this stage, can be assumed to be of the 
order of
0.1.
For t$_3$=90, f=0.1, N$_{H,24}$=0.1-1.0, and V$_3$=1.5, the derived
values for M$_{ej}$
are in the range 1.5 $\times$ 10$^{-4}$ - 1.5 $\times$ 10$^{-3}$ 
M$_{\odot}$.

Finally, the ejected mass can be estimated using the following
scaling
  law from Cassatella et al. (2005), which depends  on the approximate
assumption
that the filling factor in novae ejecta is the same as in V1668
Cyg:
\begin{equation} 
M_{ej} \sim 0.044 \cdot M_\odot/v_{exp},
\end{equation} 
where $v_{exp}$ is in km s$^{-1}$ and the  constant is set to the
values
of
the ejected mass and the expansion velocity of V1668 Cyg (Stickland
et al.
1981). Using $v_{exp}$=1500 km s$^{-1}$, one finds that for T Pyx,
 $M_{ej}$ $\sim$
2.93
$\times$ 10$^{-5}$ $M_\odot$.

We recall  that Kato and Hachisu (1991),  from their models
of steady-state winds for a nova with M$_1$=1.33 X=0.5 and Z=0.02
and the
observed t$_3$, suggested the ejection of a   massive
envelope
with M$_{ej}$ of 
about 10$^{-5}$
M$_{\odot}$ in  a single outburst.

All the previous  results agree with the considerations 
of
Shore  (1998), who
pointed out  that, for a typical ejection velocity of about -2000
kms$^{-1}$, a
nova with an optical decline time longer than a  week  must eject
a mass higher than 10$^{-5}$ M$_{\odot}$.

We also recall the fact  that in  classical novae close to maximum,
the Balmer lines develop P-Cygni profiles, which  is  a clear  
indication that a
significant  amount of material was ejected during the outburst
(Starrfield, 
1993).

Therefore, all quantitative methods and the qualitative
consideration
of  the photometric and spectral behavior of T Pyx during the
outbursts 
indicate
 the presence of a 
massive envelope with  M$_{ej}$ $\sim$ 10$^{-4}$ -10$^{-5}$ 
M$_{\odot}$.

\section{The discrepancy between the mass of the thick shell and
the  ignition mass}

The ejection of  a  massive shell  during the early (optically
thick)
outburst phase  contrast significantly  with the results of the UV +
optical
observations during quiescence  and  the theoretical
requirements
for
M$_{ign}$, which imply  a  $\dot{M}_{pre-OB}$ $\sim$ 2.2
$\times$ 10$^{-8}$ and 
a total mass for the accreted shell M$_{accr}$ of about $\sim$ 5.0
$\times$ 10$^{-7}$ M$_{\odot}$ (see Sect. 8). It is also in contrast with
the
conclusions of Sect. 9 that indicated that the closest agreement
between the
grid models and the observed properties of the system during
outburst and Q
corresponds to a model with $\dot{M}$ $\sim$ 3.0 $\times$ 10$^{-8}$
M$_{\odot}$
yr$^{-1}$ and M$_{ign}$ $\leq$ 1.3 $\times$ 10$^{-6}$ M$_{\odot}$ (Table
6).
 During outburst, T Pyx  has apparently ejected far more material
than it
has
accreted. 

Studies of classical novae containing a
massive WD indicated 
that these objects eject apparently more material than
theoretically
predicted (Starrfield et al. 1998a;  
Starrfield et al. 1998b;  Vanlandingham et al. 1996;  Shore, 1998).
These
authors emphasized the significant discrepancy 
between
the observed mass of the ejecta and the predicted critical mass of
accreted nova envelopes for massive WDs (M$_{WD}$ $\ge$ 1.25
M$_{\odot}$),
the
mass of the observed shell being one order of magnitude (or more)
higher
than that predicted by the models.

For these CNe one could attribute the discrepancy to
some
inadequacy in the TNR models or in the methods to determine the
nebular
mass. In the case of T Pyx,  the situation, however, differs because
 there is  a serious
mismatch between the shell mass indicated by the optical
{\it observations} during  outburst (M$_{ej}$ $\sim$ 10$^{-4}$
10$^{-5}$ 
M$_{\odot}$)
and that
determined 
by
the
UV and optical {\it observations} during quiescence (which give
$\dot{M}_{pre-OB}$
$\sim$
2.2 $\times$ 10$^{-8}$ M$_{\odot}$ yr$^{-1}$ and therefore 
M$_{accr}$ $\sim$ M$_{ign}$ = 5.0 $\times$ 10$^{-7}$  M$_{\odot}$). 
Therefore, the  mass ejected during outburst is about a factor of  100 
higher
than
both the theoretical ignition mass M$_{ign}$ and  the total mass
accreted 
before outburst, M$_{accr}$.

We note that using the  post 1966 value of $\dot{M}$ instead
of that
inferred for the pre-outburst interval would produce an even larger
discrepancy.

\subsection{The role of the distance}

Although  M$_{ej}$  does not depend on distance,
M$_{accr}$   does (due to its dependence on both  L$_{disk}$  and
$\dot{M}_{pre-OB}$),
and
therefore the
mismatch between M$_{ej}$ and  M$_{accr}$ depends crucially on the
assumed distance.   The uncertainty  in the
adopted distance 
was found in Sect. 2 to be of the order of 10 percent. 
However, we note,  that even the  adoption of an unlikely
distance 
of, say, 10,000 pc  (at about 20 $\sigma$ from the estimated
value)
would   only partially alleviate this inconsistency, and at the expense 
of  an uncomfortably high mass accretion rate, well
within the
range 
 for the onset of steady burning. This would imply
characteristic temperatures and 
luminosities 
that are not observed (see also Sect. 14 for a further
discussion). A
larger distance
 would also necessarily 
imply that T Pyx was super-Eddington at maximum, a
circumstance that appears unlikely due to  its slow photometric  and
spectroscopic developments during
outburst.

A lower
distance would correspond to a lower values of  L$_{disk}$, and hence
$\dot{M}_{pre-OB}$ and
M$_{accr}$,  to  values that are theoretically incompatible with
the
occurrence
of outbursts with an average interval of 22 years. It would also
exacerbate
the discrepancy between the  low value of M$_{accr}$ obtained
from
the UV observations (and the models) and the apparently high mass of
the
ejecta, as inferred from the 
behavior during outburst. Therefore, there is not much leeway to
invoke a 
different distance to explain the discordance.

\section{ The nebula revisited}

The nebula  surrounding T Pyx has been the target of several
spectroscopic and imaging observations. Duerbeck und Seitter (1979)
first
reported the presence of a  strong nebulosity
around T Pyx, with radius r $\sim$ 5", whose origin was tentatively
attributed to the 1966 outburst and whose strength was described as 
unusual. By
the assumption of an outburst expansion velocity of -900
kms$^{-1}$, a
 low
distance ($\sim$ 600 pcs) was derived.

Williams (1982) obtained spectral scans of the northern portion of
the
nebula of T Pyx. The spectrum was similar to that of  as a typical
PN,
probably photoionized by radiation from the hot remnant, and lacked
the
strong CNO enhancements characteristic of the ejecta of classical
novae.

\noindent
 Comparing images acquired  in 1979.0 and 1982.9,  Seitter (1987) found
that the
nebula did not increase in size during that time interval.

\noindent
Shara et al. (1989) from deep narrowband CCD images confirmed the
faint, 
extended
H$_{\alpha}$ + NII halo (twice as
large as the inner nebula), first reported by Duerbeck (1987).  A  
smooth,
small [OIII] nebula with r $\sim$
2" was also found. Shara et al.  (1989) also compared the relative
sizes of
the
main nebula with
r $\sim$ 5" in 1985 and in 1979 but failed as well to find any
detectable
expansion
during the 6 year interval,  confirming the finding of
Seitter
(1987).
\noindent
High resolution imagery data from HST was
obtained by Shara et al. (1997). The nebula was resolved into
more
than two thousand individual  knots, and a comparison between
images taken
at four  epochs indicated  that these individual knots retained a
similar pattern, without any evidence of expansion. These data
confirm the apparent stationarity in the 10" diameter nebula
suggested by
previous
observations. Shara et al. (1997) found  an upper limit of
40$\cdot$(1500/d)
(km s$^{-1}$) for the expansion velocity of the knots. They also
detected
nine
distinct peaks in the brightness distribution, an indication that a
multiple nebula model was required.

Many studies,
disappointingly inconclusive,  addressed the problem of the
mass of
the  nebula. In this respect, it should be noted that
large uncertainties are generally associated with  estimates of
the
mass of novae ejecta, since the mass estimate depends critically
 on
quantities that are not reliably measured, e.g.: distance, 
electron
density,  ionization structure in the nebula,  geometry, and 
filling
factor. Therefore, it is  unsurprising that a range of masses
for the nebula of T Pyx has been proposed in the
literature.

 From the nebular H$_{\beta}$ intensity measured by
Williams
(1982), WLTO (1987) 
obtained a lower limit to M$_{neb}$ of 10$^{-6}$ M$_{\odot}$, while
Shara
(1989),
using the H$_{\beta}$ 
 intensity with the requirement $\epsilon$ $\le$1 for the
filling
factor derived an upper limit of 1.0 $\times$ 10$^{-4}$ M$_{\odot}$.
From the intensity of the H$_{\alpha}$ and [NII] lines, Seitter
(1987) found a mass
close
to 8.0 $\times$ 10$^{-5}$ M$_{\odot}$. From the  H$_{\alpha}$ 
flux and 
considerations based on  HST imagery, Shara et al. (1997) obtained 1.3 
$\times$ 10$^{-6}$
M$_{\odot}$ to be the most reliable estimate for the nebula mass (with an
assumed distance of 1500
pc),
the electron density being, allegedly, the main uncertainty factor.

We add  one more  estimate for the mass of the nebula, based on
the H$_{\beta}$ flux obtained by Williams (1982) scaled to  the
entire
nebula,
after correction for the new reddening and  distance.
We obtain :

\noindent
F$_{\beta}$  $\sim$ 4.24 $\times$ 10$^{-14}$  erg cm$^{-2}$ s$^{-1}$
 and L$_{\beta}$ = 6.22 $\times$ 10$^{31}$  erg s$^{-1}$.

\noindent
 with a relative error of about 25 percent, arising from the
uncertainties in
 the distance and the flux.  
By combining  the two common relations:
\begin{equation}
L_{\beta} = 1.24\;\times \; 10^{-25} \; N_e\cdot N^+ \cdot (4/3)\;\pi\;
R_{neb}^3
\cdot \epsilon 
\label{eq:ellebeta}
\end{equation} 
and 
\begin{equation}
M_{neb}=\mu \cdot  N^+ \cdot m_H  \cdot (4/3) \; \pi\; R_{neb}^3
\cdot
\epsilon,
\end{equation}
(where $\mu$=1.4 is the mean atomic weight and    $\epsilon$  is
the
filling factor), one finds  that the mass of the nebula
(independent of 
$\epsilon$ and $R_{neb}$) is 

\begin{equation} 
M_{neb}= 18.67\;  {L_{\beta} \over  N_e} \; g = {0.584 \over N_e}
\;M_{\odot}.
\end{equation} 

\noindent
N$_e$ is inaccurately determined because, as noted by Williams
(1982),
the spectrum is too poor in number of lines to  constrain
the
physical parameters of the nebula accurately.  However, from the presence
of a
quite
strong emission feature near 3722 \AA, attributable to [OII]
3726.1 and
3728.8  one can  estimate  that the nebula density  is less
than
3.0 $\times$ 10$^{3}$  cm$^{-3}$, since these   two forbidden   lines have
critical density  log
N$_e^{crit}$ = 3.5 and 2.8, respectively. If we assume that N$_e$ $\sim$
10$^{3}$  cm$^{-3}$,  a value 
also adopted by Shara et al. (1997) from other considerations, 
we  find  that $M_{neb}$ $\sim$ 5.84 $\times$ 10$^{-4}$
M$_{\odot}$.

A value of this  order  for the
total
mass of
the nebula, which has apparently  increased from the contribution of
several
successive outbursts, agrees with the ejection during
outburst of
a massive shell ($\sim$ 10$^{-5}$ M$_{\odot}$)  as suggested
in Sect.
10.2. 

The fact   that  the  nebula of T Pyx is still clearly
observed
despite its
large distance seems hardly compatible with a  mass of
10$^{-6}$
M$_{\odot}$  derived in previous studies. In the most well
studied  CNe,
which on  average
are closer in distance than T Pyx,   
the ejected nebula has, at best, a similar strength and/or is
barely
evident a few years after outburst. 
 We think that the peculiar strength of the nebula of T Pyx
may be  explained by the fact that most of the gas ejected in
successive
outbursts has
accumulated into a  nearly stationary envelope that is strongly
irradiated by a more than average luminous UV central source.

Alternatively, one can guess that the observed nebula of T Pyx was
produced in a peculiar event and is not associated with the
recorded and/or
previous outbursts. Its 
apparent stationarity and the lack of  changes over a timescale of
about
fifteen  years (Shara et al., 1997) supports this
interpretation.
Williams (1982) noted
that the spectrum of the nebula around T Pyx is almost similar to 
typical planetary nebula, of approximately  solar composition.

We emphasize that the optically thick shell discussed in
Sects.
10.1
and 10.2 was observed only
spectroscopically during the outburst phases, and that there is no
definite,
direct
link  with the extended nebula. Simple calculations indicate that, if
d=3500 pc, 
the angular radius 
of the shell,  after ten years  of   constant expansion at
1500 km
s$^{-1}$,
would be  less than 1".  

\subsection{ The filling factor}

If the radius of the nebula  $R_{neb}$ is known,   from the
observed 
angular radius r ($\leq$ 10")  and the  distance,   one can
obtain the filling factor $\epsilon$ with the help of  eq.
\ref{eq:ellebeta}. 
 After insertion of the
values for
L$_{\beta}$ = 6.22 $\times$ 10$^{31}$ erg s$^{-1}$, R$_{neb}$ $\sim$ 5.2
$\times$ 10$^{17}$ cm, 
and  N$_e$ $\sim$ 1.1 N$^+$, one obtains that

\noindent
$\epsilon$ $\cdot$ N$_e^2$ $\sim$ 1062.7.

\noindent
Therefore, for   N$_e$ $\sim$10$^{3}$ cm$^{-3}$, 
$\epsilon$ is close to 10$^{-3}$. 
A filling factor of $\sim$ 10$^{-3}$ 
can be obtained independently from the relation of Harrison and
Stringfellow   (1994):
\begin{equation}
\epsilon \sim N_{cl} V_{cl}/V_{neb}= 8 N_{cl}^{-1/2}
\gamma^{3/2}, 
\end{equation}
\noindent
where $\gamma$ is the fraction of the spherical nebula intercepted
by the
clumps.
From the figures of Shara et al. (1997), we estimate that $\gamma$
$\sim$
0.05 
and, if
the
number of clumps   N$_{cl}$ is  $\sim$ 2000 (Shara et al. 1997), 
we find  
that
$\epsilon$
$\sim$ 2 $\times$ 10$^{-3}$.  We note that the filling factors estimated
for  
the ejecta of other novae have a wide range of values, from  4
$\times$ 10$^{-1}$ (Vanlandingham et al., 1999 for Nova LMC 1990 no.1), to
10$^{-6}$ (Saizar and Ferland (1994) for  nova  QU Vul. Williams et al.
(1981) found an intermediate value, $\epsilon$ $\sim$ 2 $\times$ 10$^{-3}$
for the
recurrent nova U Sco.  Mason et al. (2005) found  $\epsilon$ $\sim$ 
10$^{-4}$-10$^{-1}$ for nova SMC 2001 and nova LMC 2002, and  Balman and
Oegelman (1999) estimated  $\epsilon$ $\sim$   5 $\times$ 10$^{-3}$-3
$\times$ 10$^{-1}$ for
the shell of GK Per.  Mason et al. (2005) have
suggested  $\epsilon$
$\sim$
10$^{-5}$-10$^{-2}$
for T
Pyx.

\section{The mass balance and the SNIa connection}

It is  accepted that Type Ia supernovae represent the
complete
thermonuclear disruption of mass-accreting white dwarfs that reach
the
Chandrasekhar limit by accretion (Nomoto et al. 1984; Woosley and
Weaver
1986). Within this general framework,  there exist single degenerate
models
(Whelan and Iben, 1973) in which a WD accretes from a
non-degenerate
companion, and double degenerate models (e.g. Iben and Tutukov,
1984) that
involve the merger between two WDs. Hachisu (2002, 2003) 
and 
Hachisu and Kato (2001, 2002)  proposed a unified picture of binary
evolution to
SNe Ia
in which recurrent novae could be understood to be part of the 
evolutionary stages of  supersoft X-ray sources and symbiotic
channels
to SNe Ia.

The mass
M$_1$ must be  close to the Chandrasekhar limit before a recurrent nova can 
become a SN Ia, and the  WD must also increase 
in the long term, after many cycles of accretion and ejection  
(Starrfield
et al., 1985).
However, in the case of T Pyx, even if M$_1$ were  close to the
limit, which is not
clearly
established since M$_1$ appears to be close to 1.37 M$_{\odot}$,  the 
results of  the
previous sections  indicate that the mass
balance situation is unclear:

On the one hand,   the photometric and
spectroscopic behavior close to  outburst, as mentioned in Sect. 10.2, 
appear to be consistent with
 the ejection of a rather massive shell, while the UV data and theoretical 
models  (at the specific M$_1$ and $\dot{M}$
values)
indicate
that
the ignition mass 
 is   low ($\sim$5.0 $\times$ 10$^{-7}$
$M_{\odot}$). This indicates that during outburst the WD ejects 
more material than it  accumulates and that a secular decrease
in
the mass of the white dwarf is expected. Therefore, evolution to become  a
SNIa appears to be 
excluded.

On the other hand,  the ejection of a
more-massive-than-accreted shell is apparently  in 
 contrast with the  observational evidence that the chemical
composition of the T Pyx  nebula  is close to solar (Williams,
1982).  This appears to  exclude any erosion of
the white dwarf and implies that the white dwarf does not lose mass after
cycles of
accretion and ejection. However,  this  presupposes that the 
chemical composition of the
observed  nebula  is representative of
that of a single  shell ejected during outburst.

It is unclear whether these substantial discrepancies originates in
flaws
in the theoretical assumptions or the interpretation of the
observations. They certainly  highlight the  need
for
accurate
values of the most critical parameters of this recurrent nova, i.e.
the mass
and chemical composition of the shell ejected during outburst.

In any case, the behavior of  T Pyx raises several doubts about the
 common  
assumption  that in recurrent novae  far less material is
ejected during the outburst than is accreted by the white dwarf,
and
therefore that its mass increases toward the Chandrasekhar limit  
(Starrfield, 2002).
In assessing whether recurrent novae could be progenitors of Type Ia
supernovae, we note that  Hachisu and Kato (2001),   discovered that, 
among six recurrent novae, only T Pyx is offset significantly from the 
region occupied in an a orbital period - donor mass plane (their Fig. 3) by 
the progenitors of SNe Ia.

\section {Neither  a supersoft X-ray source,  nor  assisted
suicide }

To explain the alleged extremely blue color of T Pyx in
quiescence,
WLTO (1987) proposed that  nuclear burning continues even
during its
Q state, consistent with the  slow  outburst development,
which
suggests that the
accreted envelope was only weakly degenerate at the onset of TNR. 
Patterson et al. (1998)  attributed the  luminosity of  T Pyx (and V
Sagittae) to quasi-steady  thermonuclear burning and suggested  the
object to be included in the class of the supersoft X-ray sources. 

However, the color of T Pyx (B-V)$_o$ = -0.26 used in these studies
is 
based on a significant overcorrection for the reddening, assumed to
be  
E$_{B-V}$=0.36, instead of the
correct
value E$_{B-V}$=0.25  (see Paper I and earlier communications, e.g. 
Gilmozzi et al., 1998). 
It is  unfortunate that both the value 
(B-V)$_o$=-0.26  of Patterson et al. (1998) and  the statement
about the ``extremely blue color" of  T Pyx was adopted widely  in the
literature (see for example Anupama 2002, and 
Parthasaraty et al. 2007). 
We  note, incidentally, that in the same paper Patterson et al.
(1998)  
 assumed 
too high a  reddening correction (E$_{B-V}$=0.33) for  V Sge; the IUE
data  suggest,  instead a value close to 0.23. 
We recall that for T Pyx the observed (B-V)
is about 0.14 $\pm$ 0.04 (WLTO 1987; Bruch and Engel 1994; Downes
et al.
1997; Schaefer 2005; see also Table 7). This
would imply that  (B-V)$_o$ $\sim$ -0.11,  close to the value (B-V)$_o$=
-0.06
given by Szkody (1994). 

 Patterson et al.(1998)  assumed that M$_v$=1.3 and after a significant
bolometric
correction (based on the assumption of an extremely 
hot
source being  present,  
a consequence of the overestimate of the reddening), derived a
quiescent
bolometric luminosity higher  than 10$^{36}$ erg s$^{-1}$, which considered
to be a true
lower limit. This encouraged  them to invoke nuclear burning on the surface
of the
WD as the main power source,  considering the disturbingly
high $\dot{M}$
( $\ge$
10$^{-7}$ M$_{\odot}$ yr$^{-1}$) required in the case of pure
accretion
power.

\begin{figure}
\centering
\resizebox{\hsize}{!}{\includegraphics[angle=-90
]{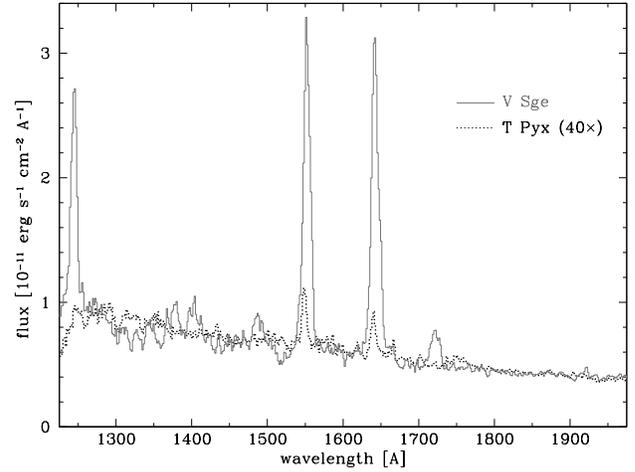}}
\caption{Comparison between the reddening-corrected  SWP 
spectra  of
the  supersoft
source
 V Sge (continuous line) and of T Pyx
(dashed line). The continuum   of T Pyx has been scaled to that  
of V Sge  (multiplied  by a factor 40).  Note the prominence of the
high
ionization lines of    NV
$\lambda$ 1240, CIV $\lambda$ 1550,  HeII $\lambda$ 1640, and  NIV
$\lambda$  1719 in V Sge,  in contrast with  their
moderate intensity  or absence in T Pyx.}

\end{figure}

However, the  IUE  and the optical data do not appear to be reproduced by
the 
model
depicted by Patterson et al. (1998),  since  the following 
observational
evidence contradicts with  their conclusions:  
\noindent
\begin{enumerate}
\item  
One of the   main results of Paper I was that  the de-reddened
UVOIR
continuum of T Pyx is reproduced well 
by a single power-law  F$_{\lambda}$ $\propto$  $\lambda$ 
$^{\alpha}$ with
a
slope $\alpha = -2.33$, representative of a
steady accretion disk.  Alternatively, as shown in Paper 1, the UV
continuum can be well fitted
by a  blackbody of  
temperature 34000 K. There is no way to reconcile these firm
observational results with the presence of a supersoft source of
typical
T $\sim$ 3.0
$\times$ 10$^{5}$ K, since its expected slope in the UV region ($\alpha =
-3.80$)
would be
inconsistent with the
UV observations. 
\item 
As shown in  Sect. 5, the   IUE observations in  1980-1996
and the optical and IR photometric data 
indicate that 
L$_{disk}$=2.7 $\times$ 10$^{35}$ erg s$^{-1}$, which corresponds to
emission  mainly 
in the directly observed UV range. 
Therefore,  in the absence of any direct  or indirect evidence of a
hot
source,  it is unlikely that L$_{disk}$ is higher  than 10$^{36}$
erg
s$^{-1}$. 
\item 
Under steady-state hydrogen-burning conditions, the accretion rate  
$\dot{M}$$_{steady}$  can be estimated to be:

\noindent
\begin{equation}
\dot{M}_{steady} \sim 3.7\:\times\: 10^{-7} (M_1-0.4) M_{\odot}\:yr^{-1},
\end{equation}

\noindent
(Hachisu and Kato  (2001), which is valid for hydrogen content X=0.7
In the case of  T Pyx, one calculates  $\dot{M}$$_{steady}$  $\sim$ 3.2
$\times$ 10$^{-7}$
M$_{\odot}$$yr^{-1}$. This
value is
almost  a factor of 30 higher than the value of $\dot{M}$ $\sim$ 1.1
$\times$ 10$^{-8}$
M$_{\odot}$  $yr^{-1}$ obtained from the  IUE data for 
the  post outburst phase. 
 \item 
   After
correction
for
inclination, the  apparent absolute  magnitude 
of M$_v$=1.79
corresponds
to 
a 4-$\pi$ averaged absolute
magnitude of M$_v^{corr}$=2.53, to be  compared with the 
average value 
M$_v^{min}$ $\sim$ 4.0 for ex-novae (Warner, 1995).
This implies that T Pyx is  more luminous than other ex-novae,  
as mentioned in Sect. 4, but is not ``extremely bright".
We can explain why T Pyx is  brighter than other novae  
in terms of  repeated   nova  eruptions  and heating of the
primary, which triggers irradiation of the secondary and produces  a
higher than average $\dot{M}$.

\item
 The emission line spectrum of T Pyx is not that of an high
excitation
object: NV $\lambda$ 1240 is nearly absent, HeII $\lambda$ 1640 is 
weaker
than CIV
$\lambda$ 1550 and barely present in
several spectra, the OIV lines close to $\lambda$ 1405  are absent. A
comparison
between the IUE spectra of T Pyx and V Sge (Fig.3)  clearly shows 
remarkable differences  and a much lower excitation  character in T
Pyx.
From an
inspection of IUE spectra of several CVs, we also found that the
spectrum
of T Pyx is  similar to that of the old novae V533 Her and
V603 Aql
and to some
spectra of the intermediate polar TV Col. The old nova RR Pic
definitely
shows higher excitation than T Pyx, with far  stronger NV and HeII
emissions.
\end{enumerate}

\noindent
Other findings  considerations also exclude the
H-burning-bloated-WD hypothesis :
\begin{enumerate}
\item
 The complexity of the optical photometric behavior in T Pyx 
(Shaefer et al.,1992;  Patterson et al., 1998) is difficult to  
explain
if the bulk of the 
luminosity
originates in a spherically symmetric   radiation
source associated with H-burning on top of a bloated white dwarf.
\item 
 If the majority of the gas  undergoes steady burning, it would be 
 difficult to understand how the remainder  that accumulates in
the degenerate envelope could burn explosively every 20 years. 
\item 
The outburst 
amplitude of T Pyx, $\sim$ 8.0 magnitudes, is  close to that found
for
classical
novae of similar t$_3$ and 
similar system inclination (Warner, 1995, 
2008). If,  as assumed by Patterson et al. (1998),  the luminosity
during quiescence is greater than  10$^{36}$ erg s$^{-1}$  then,
with an
outburst amplitude of about 8 magnitudes, the luminosity would 
reach 
10$^{40}$ erg s$^{-1}$, implying that  ~T Pyx is an object intermediate 
between a
nova and  a  SN Ia.   
\end{enumerate}

To support the hypothesis of steady nuclear burning Patterson et
al. (1998) 
considered  {\it all} CVs of
comparable
P$_{orb}$ and deduced  that the $\dot{M}$ in T Pyx was   a factor of 5000
higher
than in other CVs of  similar P$_{orb}$.  One should 
compare T Pyx with objects that are  similar,  that is,
with 
recent novae. 
After outburst,
the nova system remains in an excited state and   $\dot{M}$ 
increases
due to the  irradiation of the secondary.  
 Patterson et al. (1998)  correctly excluded from their Fig. 14
all old novae within 30 yr of the outburst
because of
their systematically too high
luminosity levels. Adopting   the same  line of reasoning,  T
Pyx
should also have  been excluded. In this respect,  we
note    
that the $\dot{M}$ of T Pyx is only  slightly higher than that
observed in
{\it recent} ex-novae (e.g. RR Pic, V841Oph, HR Del, etc)
(Selvelli, 2004).

Based on the conclusions of Patterson et al. (1998) of an the 
extremely
high luminosity (L$_{bol}$ far higher than 10$^{36}$ erg
s$^{-1}$),
Knigge et
al (2000) investigated in  detail the
evolution of the T Pyx system and proposed  that the system is a
wind-driven supersoft X-ray source. In this scenario, a strong,
radiation-induced wind is excited from the secondary star, and  
increases the rate of binary evolution, causing the system to destroy
itself
either by evaporation of the secondary star or in a Type Ia SN if the WD
reaches the Chandrasekhar limit. Knigge et al. (2000) therefore proposed
that either the primary, the secondary, or both stars may be committing
assisted stellar suicide

This scenario is, admittedly, highly speculative,  and  depends crucially
on
the unsubstantiated  assumption that both the temperature and  
luminosity
of T Pyx are extremely high. 
The IUE and optical data  are   instead consistent with a  more
conventional
scenario 
of accretion power, as in  other CVs, and we  confidently predict
that,
fortunately,
 any form of suicide in the near  future  is extremely unlikely.

Finally, we note   that Greiner et al. (2003) did not find T Pyx to be a
supersoft X-ray source, and that T Pyx does not appear in NASA's HEASARC
tool (a master compilation of EUV and X-ray databases).

\section {The  XMM   observations}

While this work was close to  completion, the  data for X-ray
observations of
T Pyx by XMM  became publicly available.
This prompted us to perform a preliminary analysis of the data to verify
the presence or absence of a supersoft source.

T Pyx was observed by XMM-Newton on November 10 2006. All the three
EPIC cameras were operated in Full Frame mode with the Medium
filter. The total useful exposure time after filtering for high
radiation periods was 22.1 ksec.
Optical Monitor data were taken simultaneously with the X-ray
observations. The values (see Table 7) are consistent with the
values 
given
in Paper I (IUE and optical observations)  and confirm the
stability of
the SED with time. 
The reduction of the XMM EPIC data was carried out  with SAS
version 7.1, using standard methods. T Pyx was detected as a faint
source that had  an observed EPIC-pn count-rate of 8.5 $\times$ 10$^{-3}$
cts
s$^{-1}$ and 
emission 
over the complete range 0.2-8 keV.

\noindent
 Figure 4 shows the XMM-Newton EPIC-pn spectrum of T Pyx 
compared to
a blackbody of 2.4 $\times$ 10$^5$ K and a luminosity of 1.0 $\times$
10$^{37}$ erg
s$^{-1}$.
 The blackbody spectrum was simulated assuming an exposure
time of 20
ksec, similar to 
that of the data, with two assumptions: 
  a distance of 3500 pc and  a reddening  E$_{B-V}$=0.25 (as in 
this paper), and a distance of 3000 pc and
a reddening of 0.4 (as assumed by Knigge et al., 2000).
The presence of a supersoft component with a temperature of 2.4
$\times$ 10$^5$ K
and a luminosity of 1.0 $\times$ 10$^{37}$ erg s$^{-1}$, whose existence
was
postulated by both 
Patterson et al. (1998)  and 
Knigge et al.(2000), can be definitely excluded.  Such a bright
component  would
be easily visible at soft energies (i.e. below 0.5 keV), and this
is
not the case. Any supersoft emission, if present, would be several
orders
of magnitude fainter than expected.

The data are compatible with the  presence of a relatively hard 
source
of   low luminosity, but  their low statistical quality  (227
counts
in the
range 0.2-8. keV) does not  allows us to perform a reliable  fit.
We also note that extrapolation of the observed IUE spectrum
(power-law)
into the X-ray range
would  result in an unrealistically high flux, several orders of
magnitude
higher than observed. The existence of such a high flux was already 
excluded in Paper I on the basis of the intensity of the HeII
1640~\AA\
line.

\begin{figure}
\centering
\resizebox{\hsize}{!}{\includegraphics[angle=-90]{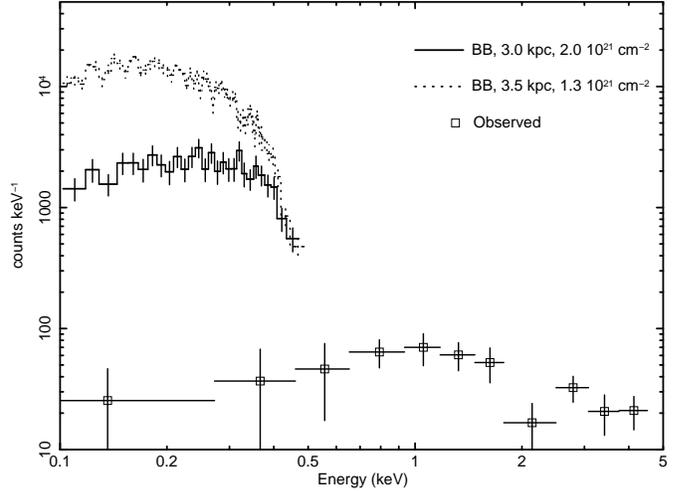}}
\caption{ The XMM-Newton EPIC-pn spectrum of T Pyx (bottom)
compared with
the
simulations of a 20 ksec exposure of a blackbody of 2.4 $\times$ 10$^5$ K
and a
luminosity of 1 $\times$ 10$^{37}$ erg s$^{-1}$  computed with  two
assumptions: a
distance of 3500 pc and  a reddening  E$_{B-V}$=0.25 (values
assumed in
this paper, dots), and a distance of 3000 pc and
a reddening of 0.4 (values assumed by Knigge et al., 2000,
continuous
line).
The three spectra shown here have been re-binned to 20 counts per
bin.}
 
\end{figure}

\noindent
We therefore ask from which process or in which source the observed 
X-ray emission originates ?
\noindent
In a CV (nova or quiescent nova)  X-rays may be produced by at
least
three different mechanisms or regions (see Kuulkers et al. 2006,
and 
Krautter,
2008):
\begin{enumerate}
\item
 Thermal emission from the hot white dwarf nova remnant in the late
outburst phases. The nova becomes a strong X-Ray emitter with a
soft SED.
\noindent
\item Emission from the inner accretion disk (BL) or the accretion
columns
(in
magnetic CVs). One expects to observe a  typical X-ray emission of CVs in
quiescence, with a thermal bremsstrahlung spectrum.
\noindent
\item Shocks in the circumstellar medium surrounding the
nova system  where the expanding nova shell and/or a nova wind
interact
with each other or with pre-existing CS material.
 The expected hard X-ray spectrum  originates from thermal bremsstrahlung 
with kTeff temperatures 0.2-15 keV, and with Lx $\sim$
10$^{33}$-10$^{34}$
erg s$^{-1}$
(O'Brien et al. 1994). 
\item
 The corona of an active dwarf M star companion. 
\end{enumerate}

In T Pyx, as already mentioned, the  XMM data would  exclude case
1. corresponding to a strong
SSS source. Case 2 can  also be ruled out  because, for a CV
accreting at
the quite high rates of T Pyx, the BL is optically thick and one
would
expect to observe (see Patterson and
Raymond 1985)  only  soft X-ray emission  with luminosity
comparable to
the disk
luminosity,  L$_{bl}$ $\sim$ L$_{disk}$ $\sim$  2.7 $\times$ 10$^{35}$ erg
s$^{-1}$.  
For case 4, we note that a high level of coronal heating at
values of
L$_x$/L$_{bol}$ $\sim$ 10$^{-3}$ persists for the active dMe stars, 
and the
emission
was  detected in surveys of nearby stars (see Giampapa and
Fleming,
2002),  although any detection would be impossible  at the
distance of T Pyx.  
 
We suggest that  the most likely origin of the observed hard X-ray
emission is from  
shocks within  the circumstellar envelope.  We note that in GK Per,
Balman
et al. (2006)   detected hard X-ray
emission by direct imaging with Chandra. The total X-ray spectrum
of the
nebula consists of two  thermal prominent components of emission. 
GK Per  has a large amount of CS material, which is  most likely a residual
of
a
planetary nebula phase, and the shell
remnant shows a
clumpy structure similar to  that observed in T Pyx by Shara et al.
(1997)
with
HST.  
We recall
that  the studies by Orio et al. (2001)  by Orio (2004)  of the
X-ray
emission from classical and recurrent novae 
demonstrated that 
emission from  shocked ejecta is expected to last about two 
years,
but 
may last for up to a century, if, for example, there is
pre-existing circumstellar material (as in the case of GK Per).  
Hernanz and Sala (2007) reported on X-ray observations of V4633 Sgr
performed
with XMM-Newton between 2.6 and 3.5 yr after outburst. The X-ray
spectrum
is
dominated by thermal plasma emission, which most probably originated in
the
shock heated ejecta.

Unfortunately, the   limited spatial
resolution  of the available XMM data do not enable any
spatially-resolved
study to be completed, because the  pixel size is  about 4 arcsec 
compared with 
the optical  radius of the nebula which is about 5 arcsec.

Also  the XMM observations, excluding the possibilities of continuous
burning and
the  supersoft source scenario (a massive white dwarf accreting at
high
rates), appear to exclude   T Pyx becoming a 
SN Ia  by means of the supersoft X-ray source channel
described by
Hachisu (2002, 2003).

\begin{table}
\caption{ XMM observations compared with IUE fluxes and ground
based photometry }
\begin{center}
\renewcommand{\tabcolsep}{0.2cm}
\begin{tabular}{ccccc}
\hline
Filter  &  $\lambda_{eff}$  & Observed flux  &  IUE flux& magn.\\
         &    (\AA)  &     (1 $\times$ 10$^{-14}$)   &     (1 $\times$
10$^{-14}$)&
\\
\hline
 UVM2  &    2310   &      0.92     & 0.80   &-  \\            
 UVW1  &    2910  &       0.86    & 0.92 &- \\          
 U    &     3440  &       0.70    & - &14.37 \\            
 B     &    4500  &       0.38    &- &15.59  \\            
 V    &     5430  &       0.24   &- &15.49   \\            
\hline
\end{tabular}
\end{center}
\end{table}

\section {The recurrence time and the next, long-awaited
outburst.}

As reported in Paper I, we started  an observing program in 1986
with IUE to monitor T Pyx prior to (and during) the expected
next
outburst that was supposed to occur in the late eighties of
the last
millennium.
Unfortunately, the star successfully  managed to postpone the
long-awaited outburst,   and
at the present time (2008)
has surpassed by eighteen years the longest inter-outburst interval
so
far recorded (24 yrs).

As mentioned  in Sect. 7, Schaefer (2005) 
published the results of a study of  the inter-outburst interval in
the recurrent novae  T Pyx and U Sco. From an analysis of the
available
data, 
 he found that the two
novae
are relatively bright during short inter-eruption intervals and dim
during
long intervals, suggesting that the product of the inter-eruption
interval
times the average bolometrically corrected flux is a constant.
Therefore, in the case
of T Pyx, the lack of the post-1967 outburst    is
explained
by a lower luminosity  and therefore  a lower $\dot{M}$. From the 
decline
in the observed
quiescent B magnitude in the time intervals before and after the
1966
outburst, Schaefer (2005)  also predicted
that the next outburst of T Pyx will occur around 2052.

With the help of  considerations in Sect. 8 and  the data
in Tables 3 and 5, we can further investigate this prediction.
The recurrence time can be estimated from the (theoretical)
M$_{ign}$
and  the
observed mass accretion rate. M$_{ign}$  depends mainly on M$_1$
and
$\dot{M}$,
while $\dot{M}$, for a given L$_{disk}$,  is a function also of
M$_1$
and R$_1$, which, in turn, is a function of M$_1$. Tables 3 and 5
list
M$_{ign}$, $\dot{M}$, and the recurrence time
$\tau$=M$_{ign}$/$\dot{M}$ (years) for various M$_1$ values.  
Table 5 (pre-1967 $\dot{M}$ values) clearly shows that 
the observed inter-outburst interval (22 years) corresponds to M$_1$
values
$\sim$ 1.36-1.38
M$_{\odot}$.  Table
3,  which contains post 1967-outburst $\dot{M}$ values indicates
that the 
observed interval,  which, so far, is longer
than  42 years,  corresponds to  M$_1$  $\le$ 1.38 M$_{\odot}$.
The  most likely value of   M$_1$ is therefore close to 1.37 M$_{\odot}$.

We note that a reduction by a factor two in the mass accretion rate
corresponds
to an
increase in the expected recurrence time $\tau$ by a larger factor
because,
for a
given M$_1$,   M$_{ign}$ increases   as $\dot{M}$
decreases. For the relevant values of  T Pyx, the decrease by a
factor of 2 in  
$\dot{M}$ is accompanied by an increase by a factor of about 50
percent
in  M$_{ign}$.
  For the next outburst,  therefore one expects an increase in
the inter-outburst interval $\tau$ by a factor of approximately 3.0,
to values
near
 60 years (see Table 3).  
Our prediction for  the next outburst date is therefore  around A.D. 2025.
With this
new date,
contrasted with that of Schaefer  (A.D. 2052),  we
(or at least some of us)  feel a bit more confident about the
chance
of
personally  testing   this  prediction.

However, given the uncertainties in $\dot{M}$ and  M$_{ign}$, the  
possibility of a more imminent outburst cannot be ruled out.  In this case,
X-ray
and
other
observations during the first outburst stages  will be of paramount
importance in
determining the mass ejected in a single event.

\section {Summary and conclusions}

We have accurately determined, from  UV and other
observations,  the
accretion  disk luminosity of T Pyx during both the pre- and
post-1966
inter-outburst phases.  For  M$_1$
$\sim$ 1.37  
M$_{\odot}$,  we have found  that  $\dot{M}_{pre-OB}$ $\sim$ 2.2
$\times$ 10$^{-8}$ M$_{\odot}$ 
yr$^{-1}$.
By combining the measured accretion rate with the  duration of the
inter-outburst
 phase
($\Delta$t = 22 yrs), the total accreted
mass is inferred  to be   
M$_{accr}$=$\dot{M}_{pre-OB}$ $\cdot$ $\Delta$t $\sim$ 5.2
$\times$ 10$^{-7}$ M$_{\odot}$.  This value is  in   
excellent  agreement with  the theoretical ignition mass 
(M$_{ign}$)
$\sim$ 5.0
$\times$ 10$^{-7}$M$_{\odot}$  expected for a massive white dwarf 
accreting
at 
the quoted rate. Therefore, both the time interval between 
the last two outburst  and
the absence of the awaited post-1967  outburst (due to the 
lower $\dot{M}$ 
in the post-1967 time interval)
are  explained  in a
self-consistent way.

This is the first reliable
  determination of the mass accreted prior to a nova outburst,
$M_{accr}$, owing  to the dominance of the
accretion
  disk luminosity over that of the secondary star at UV, optical and IR
wavelengths,
  as well as   good observational coverage during the 
inter-outburst
  phases.  Unfortunately, $M_{accr}$ cannot be confidently
determined in
other
  cases, such as classical novae, because of their  long
inter-outburst
  interval, nor in other recurrent novae, due to the faintness of
the
source,
  the lack of systematic UV observations, or the dominance of light
from
the
  giant companion over that from the accretion disk.

In T Pyx, the consistency between the observed M$_{accr}$ and the
theoretical 
M$_{ign}$ 
supports  the good quality of the observations and  the
reliability of
the models and 
represents  a new, direct  confirmation of the validity  of the TNR
theory, which associates a  massive white dwarf with the recurrent
nova
phenomenon.  
A detailed comparison of the observed parameters with the  
 theoretical  grids of Yaron et al. (2005) indicates  that the closest
agreement
 is obtained with a 
models of  a rather 
massive white dwarf  (M$_1$ $\sim$ 1.25-1.40 M$_{\odot}$) that
accretes at 
high
$\dot{M}$ rates ($\dot{M}$ $\sim$  10$^{-8}$-10$^{-7}$  M$_{\odot}$ 
yr$^{-1}$). However,    no
combination of the theoretical parameters can reproduce the observed values 
reliably, t$_3$ being the most difficult parameter to describe.

The literature data of the spectroscopic and photometric evolution
during
  the outbursts of T Pyx clearly indicate the occurrence of an
optically
thick
  phase that lasted about three months. This implies an ejected
mass of 
  M$_{ej}$ $\sim$ 10$^{-5}$ M$_{\odot}$ or higher, i.e.  much
higher than
  the mass of the accreted shell M$_{accr}$ $\sim$ 5.2 $\times$ 10$^{-7}$
M$_{\odot}$,
  inferred from  UV and other observations during quiescence.
Therefore,  T Pyx  ejected far more material
than
  it has accreted.

There is no way to reconcile this discrepancy given the 
small uncertainty in the value of  $\dot{M}$;  even if allowance
is
made for an uncertainty of a
factor two,  one obtains an upper limit 
 to 
M$_{accr}$ that  is  smaller by a factor of at least ten
than
the theoretical value of M$_{ej}$. 
Only for accretion rates higher than 4 $\times $10$^{-7}$ 
M$_{\odot}$ 
yr$^{-1}$ would the accreted mass  M$_{accr}$  be comparable with
the  
estimated  ejected mass M$_{ej}$.  
However, these high rates would correspond to the steady-burning
regime,
while 
our detailed discussion of Sect. 14  definitely excluded this
possibility.   
Further confirmation of our considerations can be found in the the very
recent
results of   XMM observations that exclude 
the
presence of a super-soft-source in T Pyx.

The important point is that far  more material appears to
have
  been ejected during the last outburst of T Pyx than has been
accreted by
the
  white dwarf. This raises several doubts about the common
assumption
  that  the white dwarf in recurrent novae increases in mass toward
the
  Chandrasekhar limit, and about the possible role of RNe as
progenitors of
  SNIa.  We note  that  Della Valle and Livio (1996), based on
statistical
considerations on the frequency of occurrence of RNe in M31 and
LMC,   
deduced that RNe are not a major class of progenitors of Type Ia
supernovae. The behavior of T Pyx represents  observational
confirmation
of this conclusion.

Further confirmation for other RNe is required, and this 
 highlights  the crucial need for   accurate
determinations of the most critical parameters of RNe,  i.e. the 
mass
accretion rate, 
and the mass and chemical composition of the shell ejected in a
single
outburst.

 In the case of T Pyx,  at present, useful information can be
obtained 
from
highly spatially  resolved spectrophotometry of  the nebula that 
resolves its innermost part  (associated with the last eruption), 
whose
apparent radius should be by now  larger than   1$''$. At the same
time,
spatially
resolved observations of the outer portions of the nebula   will
shed
light  on its   poorly known  chemical composition and on its
complex
velocity 
structure.

\begin{acknowledgements}
We gratefully acknowledge the  valuable conversations  about the
elusive
nature of this recurrent nova that we have had with many colleagues
in the
last   fifteen years, while waiting for its allegedly 
imminent
outburst. In
roughly chronological order we wish to thank in particular 
M. Livio, D. Prialnik, M. Shara, J. A. De Freitas-Pacheco, M.
Contini,  S.
Shore, R. E. Williams, M. Friedjung, O. Yaron,
J.Danziger, and  M.   Della Valle.
\end{acknowledgements}

\end{document}